\documentclass[reprint,raggedbottom,aps,physrev,amsfonts,amsmath,amssymb,floatfix,superscriptaddress]{revtex4-2}

\usepackage{graphicx,hyperref, lineno} 
\usepackage[all]{hypcap} 
\hypersetup{colorlinks=true,linkcolor=blue,citecolor=blue,urlcolor=blue}
\usepackage{siunitx}
\usepackage{amsmath}

\usepackage[caption=false]{subfig}
\makeatletter
\newcommand\xleftrightarrow[2][]{%
  \ext@arrow 9999{\longleftrightarrowfill@}{#1}{#2}}
\newcommand\longleftrightarrowfill@{%
  \arrowfill@\leftarrow\relbar\rightarrow}
\makeatother

\begin{document}

\title{Predicting Molecule Size Distribution in Hydrocarbon Pyrolysis\\ using Random Graph Theory}
\author{Vincent Dufour-D\'{e}cieux}
\email{vdufourd@stanford.edu}
\affiliation{Department of Materials Science and Engineering, Stanford University, Stanford, CA 94305, USA}
\author{Christopher Moakler}
\email{cmoakler@umd.edu}
\affiliation{Department of Mathematics, University of Maryland}
\author{Maria Cameron}
\email{mariakc@umd.edu}
\affiliation{Department of Mathematics, University of Maryland}
\author{Evan J. Reed}
\email{Evan Reed tragically passed away while this paper was in preparation.}
\affiliation{Department of Materials Science and Engineering, Stanford University, Stanford, CA 94305, USA}

\date{\today}

\begin{abstract}
    \textbf{Abstract}
Hydrocarbon pyrolysis is a complex process involving large numbers of chemical species and types of chemical reactions. Its quantitative description is important for planetary sciences, in particular, for understanding the processes occurring in the interior of icy planets, such as Uranus and Neptune, where small hydrocarbons are subjected to high temperature and pressure. We propose a computationally cheap methodology based on an originally developed ten-reaction model, and the configurational model from random graph theory. This methodology yields to accurate predictions for molecule size distributions for a variety of initial chemical compositions and temperatures ranging from 3200K to 5000K. Specifically, we show that the size distribution of small molecules is particularly well predicted, and the size of the largest molecule can be accurately predicted provided that it is not too large. 
    
 
\end{abstract}

\maketitle

\section{Introduction}

Complex chemical systems, in which a large number of chemical species are generated via numerous types of chemical reactions, play a key role in a variety of applications such as combustion \cite{harper2011combustionmechanism, westbrook2009combustionmechanism}, astrophysics \cite{kraus2017formation, hirai2009polymerization, ross1981diammonduranus}, or battery interfaces \cite{wang2018review}. Understanding the kinetic and thermodynamic forces involved is crucial for making accurate predictions of the evolution of these systems. However, estimating equilibrium chemical composition for such systems is extremely complicated. A common way to obtain  the equilibrium molecular composition is by running atomistic simulations \cite{cheng2012reaxffmechanismhydrocarbon, he2014reaxffmechanismhydrocarbon, yan2013reaxffmechanismhydrocarbon}. While these simulations eventually equilibrate, the equilibration typically takes a long time, which can be measured in weeks or even months depending on the conditions. Furthermore, each modification of the conditions, such as the temperature and the initial molecular composition, requires a brand-new simulation. 

 An example of a chemical system facing these challenges is the pyrolysis of hydrocarbons at high temperature and pressure \cite{QianArticle,EnzeArticle, QianArticle2016, QianArticle2019, QianArticle2020}, which we consider in this work. 
Hydrocarbon pyrolysis occurs in the interior of icy giant planets such as Neptune and Uranus, where small hydrocarbons, such as methane, are subject to high temperatures, typically on the order of several thousands kelvins, and high pressure, typically on the order of tens of gigapascals \cite{kraus2017formation, hirai2009polymerization, ross1981diammonduranus}. Small hydrocarbons react to form long carbon structures and these long structures can undergo phase transition to a solid phase, such as diamond. The conditions where diamond is formed have been studied in several experimental works \cite{kraus2017formation, lobanov2013carbon, zerr2006decomposition}. However, these experiments do not always agree with each other, even within conditions of temperature and pressure that are similar. These disagreements can be explained by the fact that the experimental set-ups are different between these experiments. However, a lack of understanding of the precise mechanism of hydrocarbon pyrolysis prevents researchers from having a clear understanding of the forces that are in play and their effect on the kinetic and thermodynamic equilibrium of this system.  Indeed, as these experiments are run in extreme conditions on small time scales, it is very challenging to obtain precise concentrations of the different structures that could be used to obtain these mechanisms.

As experimentally studying the hydrocarbon pyrolysis mechanism is intricate, many researchers have  studied it via modeling and simulation. Many widely-used combustion kinetic models \cite{wang2007high, zhou2018experimental} can make predictions for hydrocarbon pyrolysis; however, these models assume a gaseous phase whereas in the conditions of the interior of Neptune, hydrocarbons are in a liquid phase. In addition, these kinetic models only consider species with less than 20 atoms and hence are not suitable for the study of the growth of large carbon structures that occur in this system. Ancilotto et al. \cite{ancilotto1997dissociation} instead performed ab initio Molecular Dynamics (MD) simulations to observe the evolution of methane pyrolysis at high pressure and temperature. They observed some dissociation of methane below 100$\,$GPa and 5,000$\,$K and the formation of diamond above 300$\,$GPa. However, due to the high computational cost of ab initio MD simulations, the simulations were only run for 2$\,$ps with 16 CH$_4$ molecules. At these scales, it can be difficult to know if a system is at equilibrium or if it needs longer than a few picoseconds to react. It can also be challenging to observe the different reaction pathways that methane can undergo during its pyrolysis.

As an alternative to the high computational cost of ab initio MD simulations, classical MD simulations relying on force fields can be performed. Particularly, the ReaxFF force field \cite{van2001reaxff, chenoweth2008reaxff, srinivasan2015development, ashraf2017extension} is a common alternative to study hydrocarbon pyrolysis simulations in conditions close to the ones in Neptune or Uranus. Indeed, the parameters of this force field developed for combustion and pyrolysis of hydrocarbons have shown several successes in reproducing experimental Arrhenius parameters \cite{ashraf2017extension, cheng2012reaxff, chenoweth2009initiation, wang2011reactive, dontgen2015automated}, DFT-calculated energies \cite{qian2011reactive, wang2011reactive}, and experimentally-observed reaction pathways \cite{ashraf2017extension, beste2014reaxff, bharti2016reactive} in conditions of temperatures between 2,000 and 4,000$\,$K. In addition, these parameters were shown to be relevant for the study of condensed carbon phases by Orekhov et al. \cite{orekhov2020high} where they validated Ashraf et al. parameters \cite{ashraf2017extension} with DFT calculations. Although ReaxFF had several successes, several of its limitations have also been reported: for example, it failed to reproduce some ab initio energies \cite{ford2021nitromethane}, some common reactions were not observed in ReaxFF simulations \cite{bauschlicher2013comparison, bertels2020benchmarking}, and excited electronic states are not taken into account. 
However, due to the prohibitive computational cost of ab initio MD simulations and the intricacy of doing experiments, ReaxFF is a sensible alternative to study hydrocarbon pyrolysis in the conditions studied here, with reasonable time scales and system sizes. 
Unfortunately, ReaxFF simulations can still take several days to run which can make it time-consuming to know the evolution of hydrocarbon pyrolysis in different conditions.

In the works of Yang et al. \cite{QianArticle, QianArticle2016, QianArticle2019, QianArticle2020} and Chen et al. \cite{EnzeArticle}, kinetic models for hydrocarbon pyrolysis were built from ReaxFF MD simulations data. These models were able to reproduce the concentrations of small molecules as well as the bulk counts for larger structures arising in the MD simulations, but no information on the size of the larger structure was given. Another kinetic model for the growth of polycyclic aromatic compounds have been developed by Wang et al. \cite{wang2021molecular}; nonetheless this model does not study structures that could lead to diamond formation. 

In the recent work of E. Reed's group \cite{dufour2021atomic, dufour2022temperature}, a ``local" kinetic model capable of predicting the dynamics of counts of small and large carbon molecules was developed. In this model, trained on ReaxFF MD simulations, the reactions were described in terms of local changes at the reactive site. The local kinetic model was able to accurately reproduce the time series of the training and test MD data of counts for the largest hydrocarbons in the system. Remarkably, the initial composition and the temperature of the training set was different from that of the test set. Nevertheless, the local kinetic model suffers from a major drawback: full trajectories, using Kinetic Monte Carlo simulations \cite{gillespie1976general, higham2008modeling}, need to be run in order to obtain the equilibrium composition. Therefore, obtaining equilibrium molecular compositions in many different conditions would require significant time. In addition, this time would scale with the size of the system making it impractical to study larger size systems.

One broad goal of the scientific community studying hydrocarbon pyrolysis is to predict the structures and sizes of the products of hydrocarbon pyrolysis from any initial conditions. Of particular interest are the conditions of the interior of Neptune and Uranus where this process occurs continuously. As these structures are difficult to observe experimentally or in ab initio simulations, we chose to work with the structures observed in classical MD simulations run with the ReaxFF force field. Therefore, the specific goal of this work is to develop an easy-to-implement and fast-to-run model for hydrocarbon pyrolysis that predicts the equilibrium distribution observed in ReaxFF MD simulations of molecule sizes measured in the numbers of carbon atoms per
molecule for any given initial temperature within the range of 3200K to 5000K and any initial composition of hydrocarbons.

The proposed methodology consists of two major components.
\begin{itemize}
\item We introduce the \emph{ten-reaction model} that predicts the degree distribution given the temperature and the initial composition. This model accounts only for \emph{local} reactions of the form
$$
C_i - H + C_j - H\longleftrightarrow C_{i+1} - C_{j+1}+H - H,
$$
where the subscripts indicate the number of carbon-carbon bonds (the degree) that the reacting carbon atoms have. The number of the reactions, ten, is determined by the fact that the number of carbon-carbon bonds per carbon atom takes values only between 0 and 4. The model is calibrated on MD data with initial composition consisting of 64 C$_4$H$_{10}$ molecules at various temperatures.

\item We use the \emph{configuration model} from random graph theory \cite{MR1995,MR1998,NSW,Hofstad1,Hofstad2} to predict molecule size distribution given the degree distribution of a system, calculated with the ten-reaction model, or directly extracted from MD data. The approach based on generating functions introduced by Newman, Strogatz, and Watts \cite{NSW} is employed.
\end{itemize}

\begin{figure}[htb]
  \centering
  \includegraphics[width = 0.4\textwidth]{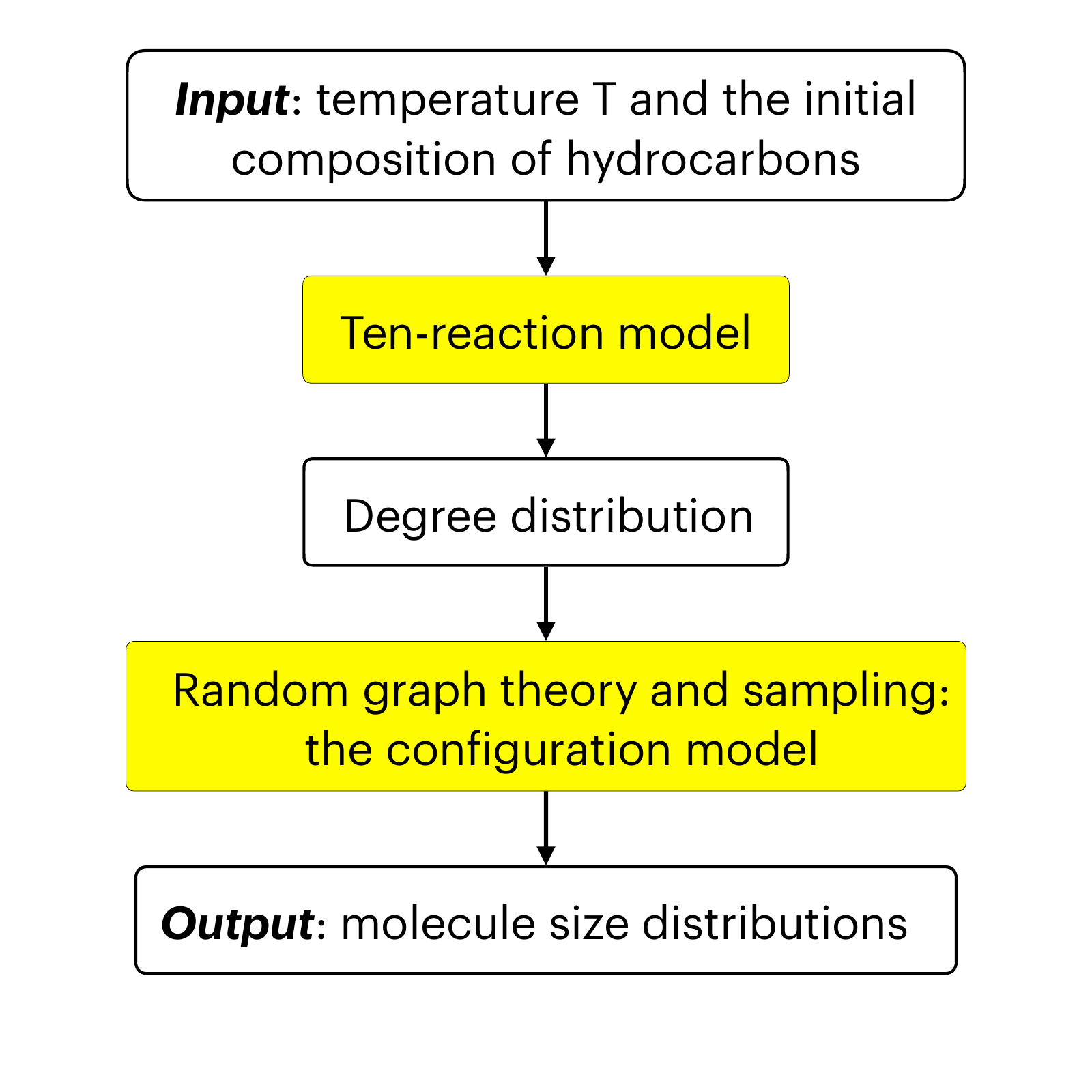}
  \caption{The flowchart of the proposed methodology.}
  \label{fig:flowchart}
\end{figure}

Knowing only the temperature and initial composition, we can use the ten-reaction model to estimate the degree distribution from which the Newman-Strogatz-Watts generating function formalism allows us to calculate the equilibrium distribution of hydrocarbon sizes we desire. The need for any ODE solver, such as kinetic Monte Carlo, is completely eliminated, as these models are analytical or based on random sampling which makes them computationally cheap. Furthermore, one can predict the histograms of small molecule counts and the histogram for the largest molecule size by sampling random graphs for the configuration model. The flowchart of the proposed methodology is shown in Fig. \ref{fig:flowchart}.

\section{Molecular Dynamics simulations for obtaining training and test data}
\label{sec:MD}
The raw data used for training and testing the model developed in this work were extracted from ReaxFF MD simulations. All these data were extracted when the simulations were at equilibrium. Equilibrium was assumed to be reached once the counts of each of the most common small molecules, CH$_4$, H$_2$, and C$_2$H$_6$, and the size of the largest molecule converged to a steady state value. These extracted data include:
\begin{enumerate}
    \item the time-averaged distribution for the numbers of carbon-carbon bonds per carbon atom, or, in other words, the time-averaged degree distribution for the graphs consisting of the union of the carbon skeletons of all molecules present at each time step;
    \item the time-averaged counts of molecules of each size, i.e., the molecule size distribution.
\end{enumerate}

The MD simulations were run using the LAMMPS software \cite{PLIMPTON19951, thompson2022lammps} and the KOKKOS package \cite{CarterEdwards20143202}. The ReaxFF potential \cite{van2001reaxff,ReaxFF_2, AKTULGA2012245} with the parameters described by Ashraf et al. \cite{ashraf2017extension} was used as the interatomic potential. The initial molecules present in the simulations were: exclusively C$_4$H$_{10}$, C$_2$H$_6$, CH$_4$, C$_8$H$_{18}$, or a mix of molecules. The numbers of molecules for each simulation were chosen to have a total number of atoms between 800 and 1,000 in each simulation. The simulations were run for a duration allowing the system to reach an equilibrium and remain in it for a sufficient period of time (see Supplementary Section \ref{sec:MD_details} and Fig. \ref{fig:eq_def} for the definition of equilibrium). The simulation was between 40 and 1000 ps. To take into account the variance between MD replicates, two MD simulations for each set of conditions were run. The simulations details, the conditions of each of the simulations, and the duration during which the simulations are at equilibrium can be found in the Supplementary Materials Section \ref{sec:MD_details} and Table \ref{table:MD}. The computation time to run the MD simulations was around 1 week on 40 CPUs for 500 ps of simulation.

An important aspect of extracting data from MD simulations is the criterion for when two atoms should be treated as bonded. We used the bond duration and bond length criterion proposed in \cite{EnzeArticle}: two atoms are considered to be bonded if the interatomic distance has been less than a threshold distance $\lambda$ for a threshold duration $\tau$. The values for $\lambda$ were taken from Chen et al. \cite{EnzeArticle}: $\lambda = 1.98\,\si{\angstrom}$ for C-C bonds, $\lambda = 1.57\,\si{\angstrom}$ for C-H bonds, $\lambda = 1.09\,\si{\angstrom}$ for H-H bonds. The bond duration was studied at different temperatures using the protocol developed by Chen et al. \cite{EnzeArticle} (see Supplementary Materials Section \ref{sec:bond_duration}) and the value $\tau = 0.096\,$ps was set for all temperatures.

\section{Methods}
\label{sec:methodology}
The proposed methodology consists of two steps. Step one predicts the degree distribution for the random graph comprised by carbon skeletons of all molecules with the aid of the originally developed \emph{ten-reaction model}. Step two calculates the molecule size distributions from the degree distribution using the generating function approach \cite{NSW} or a random graph sampling approach. Below we detail each of these steps.

\subsection{The ten-reaction model for predicting the degree distribution}
\label{sec:predict_deg_distr}

In each MD simulation, tens of thousands of chemical reactions are observed. These reactions can be described at the molecular scale, where the whole molecules involved are described, or at the atomic scale, where only atoms surrounding the reactive site are described. A comparison of these two descriptions in the case of hydrocarbon pyrolysis was performed previously \cite{dufour2021atomic}. Here, we wish to obtain the degree distribution of the carbons, which is an atomic property; therefore, we will describe reactions at the atomic scale. When observing the reactions at the atomic level, these reactions can be divided into three groups according to the change in the total number of the carbon-carbon bonds (CC-bonds) in the system:
\begin{enumerate}
    \item there is no change in the CC-bonds;
    \item a single CC-bond is added or removed;
    \item more than one CC-bonds are added or removed.
\end{enumerate}

The histogram of changes in the CC-bonds in the reactions observed in the MD simulation at $T=3300$K initialized with 64 molecules C$_4$H$_{10}$ is shown in Fig. \ref{fig:CCchangeHist}. In this case, approximately 50\% of all reactions observed did not change the total number of CC-bonds. Respectively, 24.25\% and 24.32\% of the reactions removed or added a single CC-bond. The remaining 1.4\% of the reactions added or removed two or more CC-bonds. The reactions that do not lead to changes in CC-bonds do not affect the molecule size distribution measured in the number of carbon atoms in the molecules. Over 97\% of the remaining reactions either add or remove exactly one CC-bond. These observations motivate our first simplifying assumption:

{\bf Assumption 1.} \emph{Each reaction changes exactly one CC-bond.}

As we are studying the equilibrium degree distribution, we consider only the molecules that are stable. This means that we ignore radicals as these molecules are only transition species and do not remain for a long time in this state. We also ignore double bonds. The reason for this assumption is that they are not prevalent. Except at temperatures above 4,500 K more than 90\% of carbons are bonded to 4 atoms, as can be observed in Fig. \ref{fig:PCbonds}. The rest of the carbons are mostly bonded to 3 other atoms. These carbons bonded to 3 other atoms are either radicals or forming a double bonds and their total percentages represent less than 10\% of all the carbons. Ignoring radicals and double bonds brings us to our second simplifying assumption:
 
{\bf Assumption 2.} \emph{All carbons have four single bonds}. 

Using Assumptions 1 and 2, we obtain the following set of possible reactions:

\begin{equation}
   C_i-H + C_j-H \xleftrightarrow{\text{$K_{ij}$}} C_{i + 1} - C_{j + 1} + H - H.
   \label{eq:reactions}
\end{equation}

The subscripts of the carbons indicate their degrees, i.e., $C_k$ is bonded to  $k$ other carbon atoms. Since the valence of carbon is 4, the degree $k$ can be $0$, $1$, $2$, $3$, or $4$, resulting in ten possible reactions. We note that these reactions are not elementary reactions and would likely need several elementary steps to occur. However, these reactions are considered to govern the degree distribution of the carbons at equilibrium.

Using this simplified mechanism, the equilibrium constants $K_{ij}$s can be fitted to the data obtained with the MD simulations. The $K_{ij}$s are fitted at different temperatures to MD simulations starting with C$_4$H$_{10}$. A least squares fit to the Arrhenius model is used to obtain the $K_{ij}$s as a function of temperature.

Once the parameters for the  Arrhenius models for $K_{ij}$s are obtained, the degree distribution at equilibrium can be readily calculated given the temperature and the initial chemical composition of the system. The calculation consists in solving a system of nonlinear algebraic equations. 
The details of the fitting of the equilibrium constants $K_{ij}$s, finding the degree distribution, and the uncertainty quantification for the degree distribution are provided in the Supplementary Materials Section \ref{sec:10RM_details}.

\subsection{Generating molecule size distribution from degree distribution} 
\label{sec:comp_size_distr}
The collection of carbon skeletons of hydrocarbon molecules at each step of an MD simulation can be viewed as the realization of a random graph whose vertices are the carbon atoms (C-atoms) and edges are the carbon-carbon bonds (CC-bonds). We remind the reader that a graph consists of sets of nodes and edges. In undirected graphs, each edge connects an unordered pair of nodes. The hydrogens are not represented in the graphs. The degree of a node is the number of edges that this node forms. In our case, the degree of a C-atom represents the number of CC-bonds that this carbon forms.

The degree distribution for this random graph can be extracted from the MD simulation or predicted as described in Section \ref{sec:predict_deg_distr}. Therefore, this random graph is an instance of the so-called \emph{configuration model} \cite{MR1995,MR1998,NSW} briefly described below. A detailed derivation of this model is given in Ref. \cite{NSW}. Its adaptation for the case of hydrocarbons is presented in the Supplementary Materials Section \ref{sec:RGT_details}.

Our goal is to obtain the probabilities $\pi_s$, $s \in\mathbb{N}$, that a randomly picked molecule has $s$ C-atoms, using the degree distribution $p_k$. For this purpose, we define the probabilities $P_s$, $s\in\mathbb{N}$, that a randomly chosen C-atom belongs to a molecule containing $s$ C-atoms. The distribution $\{\pi_s\}$ is readily obtained from the distribution $\{P_s\}$:
\begin{equation}
    \label{eq:pis} \pi_s = \frac{P_s/s}{\sum_{j\in\mathbb{N}} P_j/j},\quad s\in\mathbb{N}.
\end{equation}

To obtain the distribution $P_s$ from the degree distribution $p_k$, we use the generating functions of these distributions \cite{Wilf}. For any discrete probability mass function (PMF)  $\{a_k\}$, the generating function is defined by $G(x):=\sum_{k} a_kx^k$. Generating functions possess a number of useful properties.  For example, since any PMF sums up to 1, we have $G(1) = 1$. The definition of the mean implies that $\langle k\rangle = G'(1)$. We denote the generating functions of the distributions $p_k$ and $P_s$ by $G_0(x)$ and $H_0(x)$ respectively. While $G_0(x)$ can be readily written out ($G_0(x) = p_0 + p_1x+p_2x^2+p_3x^3 +p_4x^4$), finding $H_0(x)$ is nontrivial and requires more work.

To find $H_0(x)$, we need to define the so-called \emph{excess degree distribution} $\{q_k\}$ as follows. When an edge is randomly picked and then one of its endpoints $v$ is chosen,  the probability that $v$ has $k$ other edges incident to it is 
\begin{figure}[htb]
  \centering
  \includegraphics[width = 0.4\textwidth]{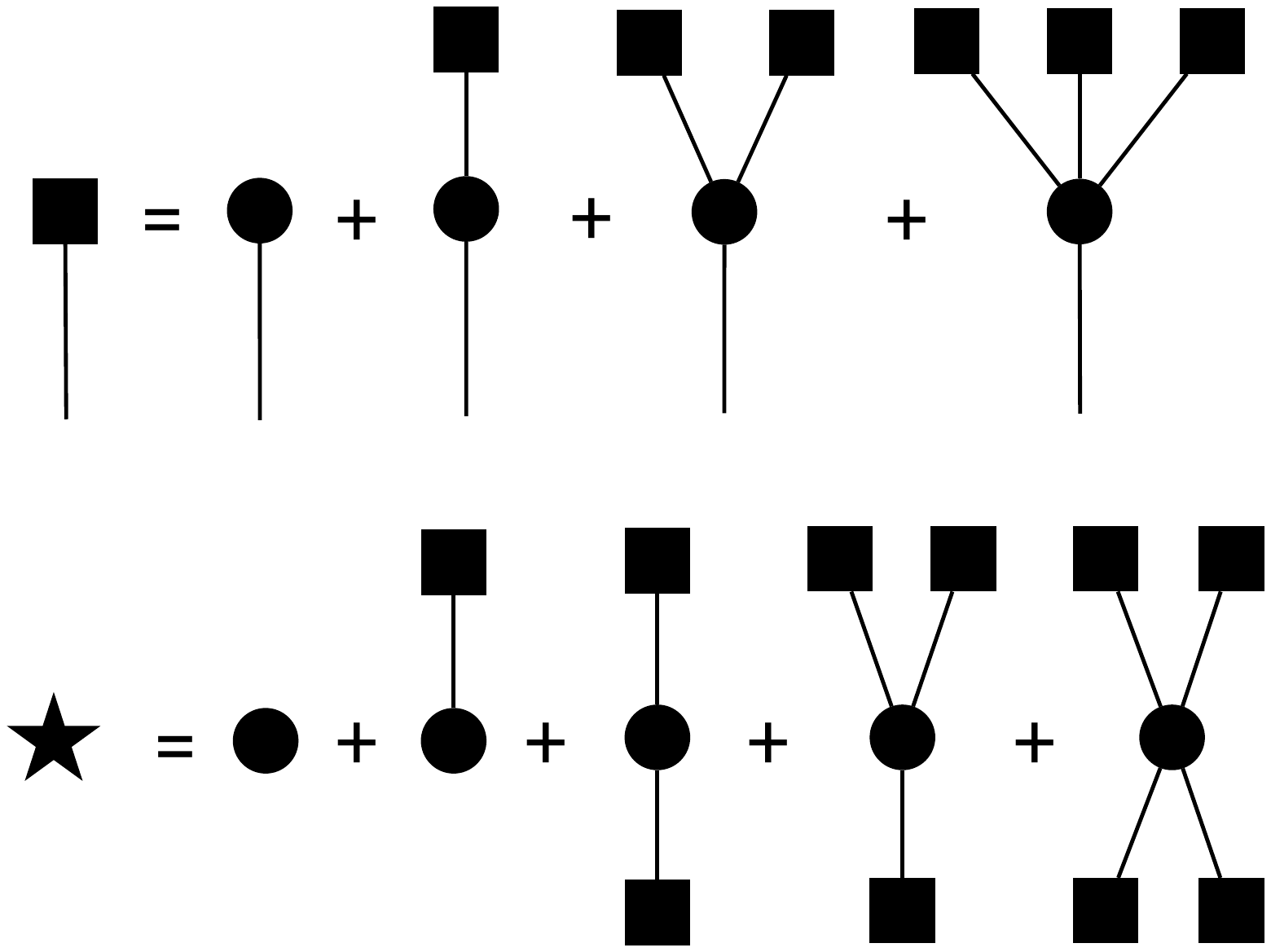}
  \caption{A graphical explanation for the self-consistency relationship \eqref{eq:H1self} (top) and equation \eqref{eq:H0} (bottom) for the case relevant to random graphs representing carbon skeletons of hydrocarbons where the functions $G_1$ and $G_0$ are cubic and quadratic polynomials respectively. The symbols $\blacksquare$ and $\bigstar$ denote the distributions $H_1$ and $H_0$ respectively. The circles represent vertices of the graph. {\it Top:} The size distribution $H_1$ of the connected component that can be reached by picking a random edge and a random direction along it is a sum of the following four cases. Let $v$ be the vertex defined by this random edge and this random direction. Case 0: the vertex $v$ has excess degree 0 with probability $q_0$. Case 1: $v$ has excess degree 1 with probability $q_1$; in this case, the size distribution of the connected component hanging off the other edge incident to $v$ is again generated by $H_1$. In Cases 2 and 3, $v$ has two and three other edges incident to it with probabilities $q_2$ and $q_3$ respectively. The size distributions of the connected components hanging off each of those edges are generated by $H_1$ as well. {\it Bottom:} The size distribution of a connected component containing a randomly picked vertex $v$ is generated by $H_0$ which adds up from the following five cases. Case 0: $v$ has degree zero with probability $p_0$. Cases 1--4: $v$ has degree $k$ with probability $p_k$, $k=1,2,3,4$. In these cases, the size distributions of the connected components hanging off each edge incident to $v$ are generated by $H_1$.}
  \label{fig:self}
\end{figure}
\begin{equation}
\label{eq:qk}
q_k = \frac{(k+1)p_{k+1}}{G_0'(1)}.
\end{equation}
The generating function $G_1(x)$ of the excess degree distribution is readily found: $G_1(x) = G_0'(x)/G_0'(1)$. Using $G_1(x)$, we can obtain the generating function $H_1(x)$ for the probability distribution for the component sizes hanging off from one of the ends of a randomly picked edge. The calculation of $H_1(x)$ stems from the following two ideas. The first idea is the use of \emph{the power property} of any generating function: if $G(x)$ generates the distribution for some quantity associated with an object, then $[G(x)]^m$ generates the distribution for the sum of these quantities of $m$ independent samples of the object. The second idea is that $H_1(x)$ must satisfy the following self-consistency relationship:
\begin{equation}
    \label{eq:H1self}
    H_1(x) = xG_1(H_1(x)).
\end{equation}
An insightful graphical explanation for \eqref{eq:H1self} was offered by Newman, Strogatz, and Watts \cite{NSW}. Fig. \ref{fig:self} (top) displays a version of it adapted for our case where the excess degree distribution has at most four nonzero components: $q_0$, $q_1$, $q_2$, and $q_3$.
Since $G_1(x)$ is a cubic polynomial, equation \eqref{eq:H1self} with respect to $H_1$ has three roots. Only one of them does not blow up as $x\rightarrow0$, allowing us to obtain a unique solution for $H_1(x)$. 

The generating function $H_0(x)$ can be obtained from $H_1(x)$ using the formula (see Fig. \ref{fig:self} (bottom)):

\begin{equation}
    \label{eq:H0}
    H_0(x)=xG_0(H_1(x)).
\end{equation}
The distribution $\{P_s\}$ can be extracted from its generating function $H_0$ using Cauchy's integral formula (Eq. \ref{eq:restore}). Then the probabilities $\pi_s$ that a randomly picked molecule has $s$ C-atoms can be obtained from $\{P_s\}$ using equation \eqref{eq:pis}. 

Two different regimes can be observed for the molecule size distribution: either all molecules remain small (subcritical regime) or one molecule, called the \emph{giant component}, contains a substantial fraction of carbons at all times (supercritical regime). These two regimes are separated by the criterion \cite{MR1995,NSW}:

\begin{equation}
\label{eq:crit}
    \sum k(k-2)p_k\begin{cases} <0,&\text{subcritical regime}\\> 0,&\text{supercritical regime}\end{cases}.
\end{equation}

The distributions $H_0$ and $H_1$ account only for small tree-like connected components of a random graph. In the subcritical regime where there is no giant component, these distributions sum up to 1. If there is a giant component, $H_0$ sums up to the fraction of vertices {\it not} lying in the giant component. Hence, when the system is in the supercritical phase, one can estimate the fraction $S$ of vertices in the giant component using the relationship \cite{NSW}:
\begin{equation}
    \label{eq:largestcompsize}
    H_0(1) = G_0(H_1(1)) = 1-S.
\end{equation}
Therefore, the expected number of carbons in the largest hydrocarbon molecule is $N_C(1-H_0(1))$, where $N_C$ is the number of carbons in the system.  Equation \eqref{eq:H1self} and Fig. \ref{fig:self} (top) imply that   $H_1(1) = u<1$ in the supercritical phase, where $u$ is the smallest positive solution to $u=G_1(u)$. 

The model described in this section will be referred to as the ``RGT'' model in the rest of this paper. RGT stands for \emph{Random Graph Theory}.

\subsection{Random graph sampling}
\label{sec:RGS}
The algorithm for obtaining the molecule size distribution $\{\pi_s\}$ and the size of the giant component (if any) summarized in Section \ref{sec:comp_size_distr} is extremely cheap: it runs in less than a second on a single CPU. However, it is based on the assumption that the number of vertices (carbons) is very large. 
An alternative way to predict molecule size distribution from the degree distribution is by sampling random graphs for the configuration model and applying standard graph algorithms such as depth-first search \cite{CLRS} to find all connected components. Random graph sampling has several advantages. First, it allows us to fix the number of vertices to match the number of carbons in the considered system and hence eliminate the assumption that the system is large. Second, it offers a direct way to approximate any distribution we want. In particular, we are interested in the size distribution for the largest molecule. This distribution depends on the size of the system and hence cannot be obtained via the generating function approach. On the downside, it takes about a minute to generate and process a large enough collection of random graph samples on a regular modern laptop. Moreover, all estimates obtained using random graph sampling unavoidably contain statistical errors.

Given the degree distribution and the total number of vertices $N_C$ (carbon atoms), we can create a collection of random graphs as outlined in \cite{NSW}. The main idea of this procedure is to generate a collection of vertices with ``stubs" (i.e. edges sticking out of them) whose number obeys the degree distribution and then randomly pair the stubs:
\begin{enumerate}
    \item The carbon atoms are enumerated from 1 to $N_C$. The first ${\sf round}(p_0N_C)$ carbons get zero stubs. The next ${\sf round}(p_1N_C)$ carbons get one stub each. The next ${\sf round}(p_2N_C)$ carbons get two stubs each, and so on.
    \item The total number of edges is one half of the number of the stubs. If the number of stubs turns out to be odd, the last atom with zero stubs is given one stub. Then the stubs are matched as follows. First, the stubs are enumerated. Next, a random permutation of the stubs is generated. Then the stubs from the top half of this permutation are paired with the stubs of its bottom half. As a result, a realization of the random graph is generated.
    \item Finally, the realization of the random graph is postprocessed by removing self-loops and repeated edges.   
\end{enumerate}

This random graph sampling model will be referred to as the ``RGS'' model in the rest of the paper.

\section{Results}
\label{sec:results}
In this section, we apply our proposed methodology to predict the molecule size distribution for hydrocarbon pyrolysis. Since our methodology involves several components, we test each of them separately to achieve a more detailed assessment of its strengths and limitations.

\subsection{Prediction of the degree distribution}
\label{sec:res_pred_degr}

\begin{figure*}[!htb]
  \centering
  \includegraphics[width=0.9\textwidth]{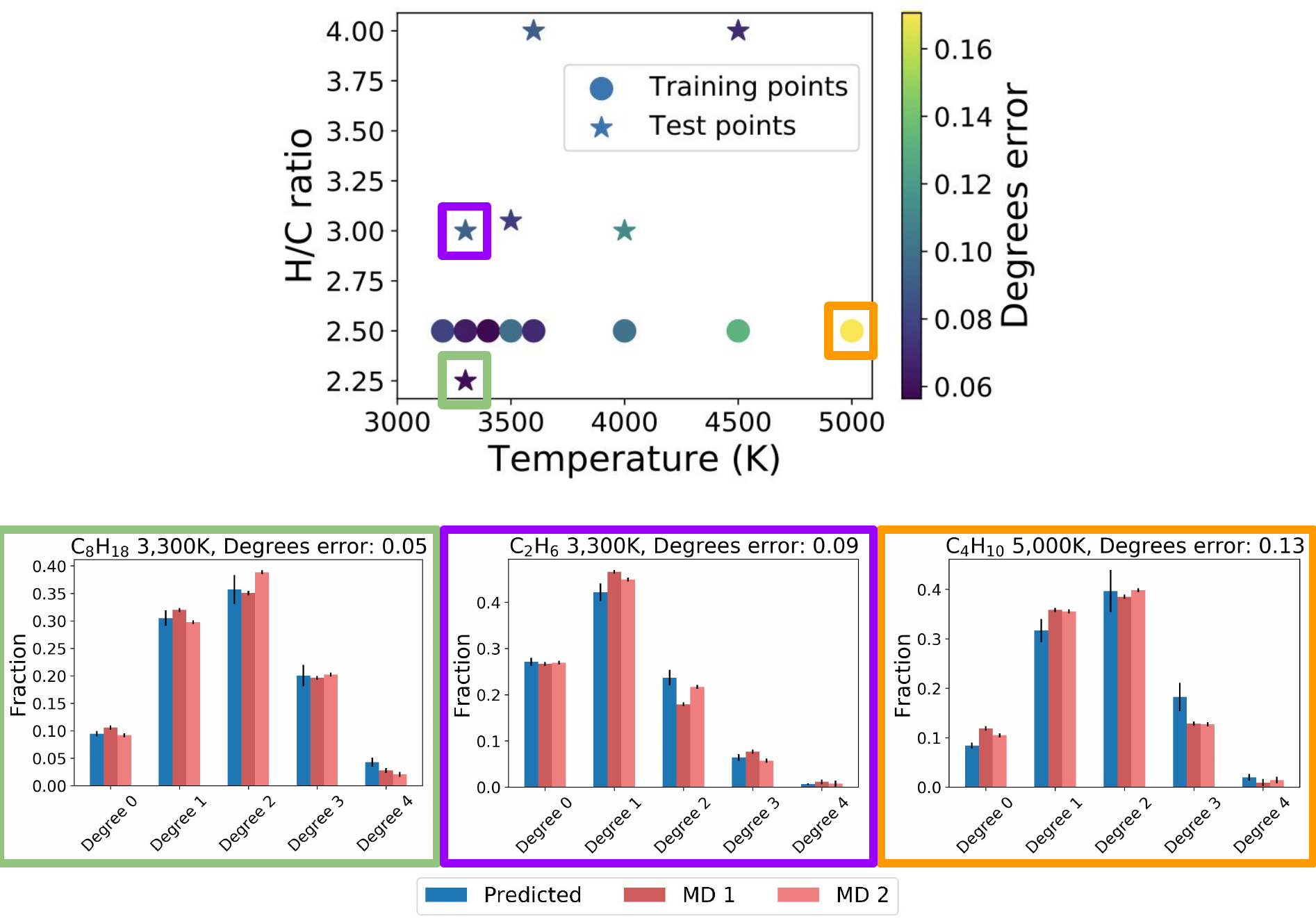}
  \caption{A comparison between the degree distributions predicted by the ten-reaction model and extracted from the MD simulations. Top: The degrees error Eq. \eqref{eq:error} as a function of temperature and the H/C ratio is showed using color coding.  Bottom: Examples of small  $E=0.05$ (left, green frame), medium  $E=0.09$ (middle, purple frame) and relatively large $E=0.13$ (right, orange frame) errors arising in three different systems. The corresponding marks in the top plot are surrounded with squares of the corresponding colors.
  }
  \label{fig:Degrees_pred}
\end{figure*}

We have assumed that the equilibrium constants $K_{ij}$ obey the Arrhenius law for any hydrocarbon composition. The validity of this assumption is demonstrated in Fig. S5 displaying Arrhenius plots for  $K_{ij}$s  for a collection of training and test data for temperatures in the range 3200K $\le T\le$ 5000K.

To compare the degree distributions predicted by the ten-reaction model and extracted from MD simulations, we define the following error metric:
 \begin{equation}
   E = \sum_{i=0}^4 \left|\left(\frac{N_{C_i}}{N_C}\right)^{\rm 10RM} - \left(\frac{N_{C_i}}{N_C}\right)^{\rm MD}\right|,
   \label{eq:error}
\end{equation}
where the superscripts 10RM and MD mean that the corresponding data come, respectively, from the ten-reaction model and from the average of two MD simulations.

Fig. \ref{fig:Degrees_pred} visualizes this comparison. It can be observed that even for the highest error, the predicted degree distribution matches well the degree distribution extracted from the MD data.

While the error in the predicted degree distribution is fairly small for all initial compositions and temperatures that we have considered, there is an apparent trend observed for C$_4$H$_{10}$ data: the error increases as the temperature grows. This trend and its consequences will be discussed in Section \ref{sec:discussion}. 

\subsection{Prediction of small molecule size distribution}
\label{sec:results_mol_size_distr}
Hydrocarbon pyrolysis usually generates a mix of small molecules, and one much larger molecule, if the H/C ratio is low enough. In order to accurately predict the equilibrium conditions of hydrocarbon pyrolysis we need to correctly describe both of these components of the system. The algorithm summarized in Section \ref{sec:comp_size_distr} predicts the size distribution for small molecules as well as the mean size of the largest carbon agglomeration, called the \emph{giant component} in the random graph theory framework. Here, we start by studying the small molecules size distribution. We chose 20 carbons in a molecule as a reasonable size cut-off for ``small" molecules. 

  For each initial composition and temperature reported in Table S1, the following three small molecule size distributions were compared:
 \begin{itemize}
     \item The molecule size distribution extracted from MD simulations.
     \item  The molecule size distribution predicted by the configuration model described in Section \ref{sec:comp_size_distr} using the degree distribution extracted from the MD simulations (``RGT'' model).
     \item The molecule size distribution predicted by the configuration model described in Section \ref{sec:comp_size_distr} using the degree distribution predicted by the ten-reaction model described in Section \ref{sec:predict_deg_distr} (``10RM + RGT'' model).
 \end{itemize}
 
\begin{figure*}[!htb]
  \centerline{
\includegraphics[width=\textwidth]{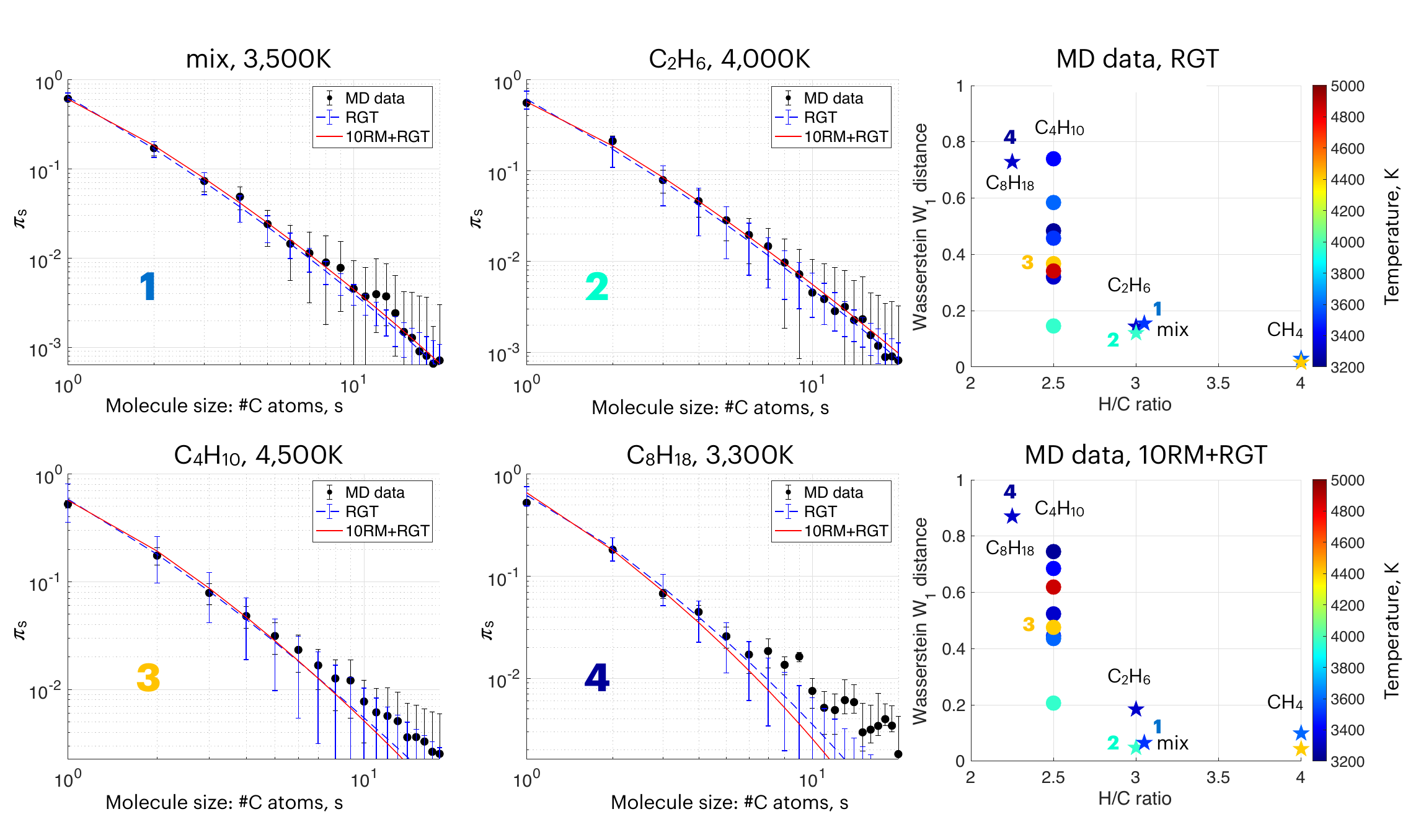}
 }
  \caption{{\it Left and Middle Columns}: The predicted molecule size distributions for hydrocarbons with at most 20 carbon atoms are compared to the ones extracted from MD simulations (solid black dots). The black error bars are equal to two standard deviations. The molecule size is measured by the numbers of carbon atoms $s$. The dashed blue curves and the solid red curves correspond to predicted size distribution by the random graph theory model (RGT). The dashed blue curves with error bars  represent the results obtained using the degree distributions extracted from the MD simulations, while the red solid curves represent the results obtained using the degree distribution generated using the ten-reaction model (10RM). {\it Right Column}: The Wasserstein-1 distance $W_1(P,Q)$, where $P$ is the molecule size distribution extracted from MD simulation and $Q$ is the predicted molecule size distribution by the ``RGT'' model (left) and the ``10RM + RGT'' model (right) plotted as a function of the H/C ratio. The initial composition is indicated for each mark. The temperature is shown with color coding.}
  \label{fig:MSDistr}
\end{figure*}

    Fig. \ref{fig:MSDistr} (left and middle columns) illustrates four samples of predicted and extracted from MD simulations molecule size distributions. These samples are chosen representatively with respect to initial compositions, temperatures, and discrepancies between the distributions extracted from MD simulations and the ones calculated via the random graph theory. More plots of this kind are found in Fig. \ref{fig:small_mol_fig} in Supplementary Materials. We observe that the predictions of the RGT and 10RM + RGT models match generally well the values observed in the MD simulations, especially for the smallest molecules (containing less than 5 carbons). We can also note that the RGT and 10RM + RGT model predictions are very close to each other. 
 
In order to compare the accuracy of the prediction for the small molecule size distribution for different temperatures and initial chemical compositions, we use the Wasserstein $W_1$ distance,  a standard measure for comparing probability distributions:

\begin{equation}
    \label{eq:W1}
    W_1(P,Q) = \sum_{s}\left|\sum_{j=1}^sP_j - \sum_{j=1}^s Q_j\right|.
\end{equation}
The molecule size distributions extracted from MD simulations have been used as the distribution $P$, while the predicted molecule size distributions have been used as $Q$. In both cases, we truncated these distributions at size $s=20$ and renormalized them so that they sum up to 1. The values of the Wasserstein $W_1$ distance obtained for various initial compositions and various temperatures are visualized in Fig. \ref{fig:MSDistr} (right column). This figure allows us to make the following observations.

\begin{itemize}
    \item The Wasserstein $W_1$ distance tends to increase as the H/C ratio decreases. This phenomenon is related to the presence of large hydrocarbons. We will discuss it in Section \ref{sec:discussion}.
    \item 
    The temperature dependence observed for C$_4$H$_{10}$ data is nonmonotone.
    \item The difference between the accuracy of RGT and 10RM+RGT for small molecule size distributions is insignificant.

\end{itemize}

 The RGT model is derived under the assumption that the number of vertices in the graph is very large. Hence we expect that the predictions by RGT will become more accurate as the number of carbons in the system increases. This effect is illustrated in Section \ref{sec:size_effect} of Supplementary Materials.

The RGS model can also be used to study the small molecule size distribution. In particular, this method can be used to obtain a histogram of the counts of molecules of a particular size at equilibrium. More details and figures are given in Section \ref{sec:hist_small} of Supplementary Materials.

\subsection{Prediction for the size of the largest molecule}
\label{sec:hist_largest}

There are two ways to predict the size of the largest molecule given the degree distribution. If the degree distribution implies the supercritical phase and the number of carbons is large enough, the size of the giant component can be estimated by $N_C(1-H_0(1))$ (RGT, see the end of Section \ref{sec:comp_size_distr}). Alternatively, for a system of any size and independent of phase, the size distribution of the largest molecule can be obtained using random graph sampling (RGS) (see Section \ref{sec:RGS}). The results of both of these approaches are presented here.  The degree distribution is obtained by the ten-reaction model (10RM).

The existence of the giant component was predicted for initial compositions consisting of C$_4$H$_{10}$ molecules and C$_8$H$_{18}$  molecules, while for the mix and the initial compositions consisting of C$_2$H$_6$ and CH$_4$, the fraction of carbons in the giant component $S$ was found to be zero. We report the numbers of carbons in the giant component $S\cdot N_C$ in Table \ref{table:giant}. We can see that the size of the giant component is generally over-estimated by the RGT model. This observation will be discussed in Section \ref{sec:discussion}.

\begin{table}[!htb]
    \centering
    \begin{tabular}{c|c|c|c|c|c}
        Init. Comp. &Temp.&$N_C$&RGT&RGS&MD\\
        \hline
        C$_4$H$_{10}$ & 3200K & 256 & 79 & $112 \pm 25$ & $57\pm 18$\\
        C$_4$H$_{10}$ & 3300K & 256 & 81 & $110 \pm 26$ & $49\pm 17$\\
        C$_4$H$_{10}$ & 3400K & 256 & 115 & $113 \pm 26$  & $70\pm 28$\\
        C$_4$H$_{10}$ & 3500K & 256 & 112 & $113 \pm 27$ & $64\pm 21$\\
        C$_4$H$_{10}$ & 3600K & 256 & 125 & $117 \pm 26$ & $76\pm 24$\\
        C$_4$H$_{10}$ & 4000K & 256 & 92 & $129 \pm 25$ & $57\pm 22$\\
        C$_4$H$_{10}$ & 4500K & 256 & 105 & $150 \pm 22$ & $58\pm 21$\\
        C$_4$H$_{10}$ & 5000K & 256 & 96 & $171 \pm 18$ & $54\pm 19$\\
        C$_8$H$_{18}$ & 3300K & 288 & 195 & $208\pm13$  & $127\pm 36$\\
        C$_2$H$_6$ & 3300K & 250 &-- & $24\pm 8$& $20\pm 6$\\
        C$_2$H$_6$ & 3300K & 250 &-- & $31\pm 11$& $20\pm 5$\\
        CH$_4$ & 3600K & 160 &-- & $6\pm 1$ & $6\pm 1$\\        
        CH$_4$ & 4500K & 160 &-- & $10\pm 3$& $10\pm 3$\\
        mix & 3500K & 240 &-- & $21 \pm 7$& $18\pm 3$\\
    \end{tabular}
    \caption{The numbers of carbon atoms in the largest molecule. The column RGT shows the predictions by 10RM+RGT: the expected number of carbons in the giant component is the product of $S$ and the number of carbons $N_C$. The last two columns display {\tt mean} $\pm$ {\tt standard deviation} for the size of the largest molecule extracted from generated by 10RM+RGS and MD simulations respectively.} 
    \label{table:giant}
\end{table}

\begin{figure*}[!htb]

  \centerline{
\includegraphics[width=\textwidth]{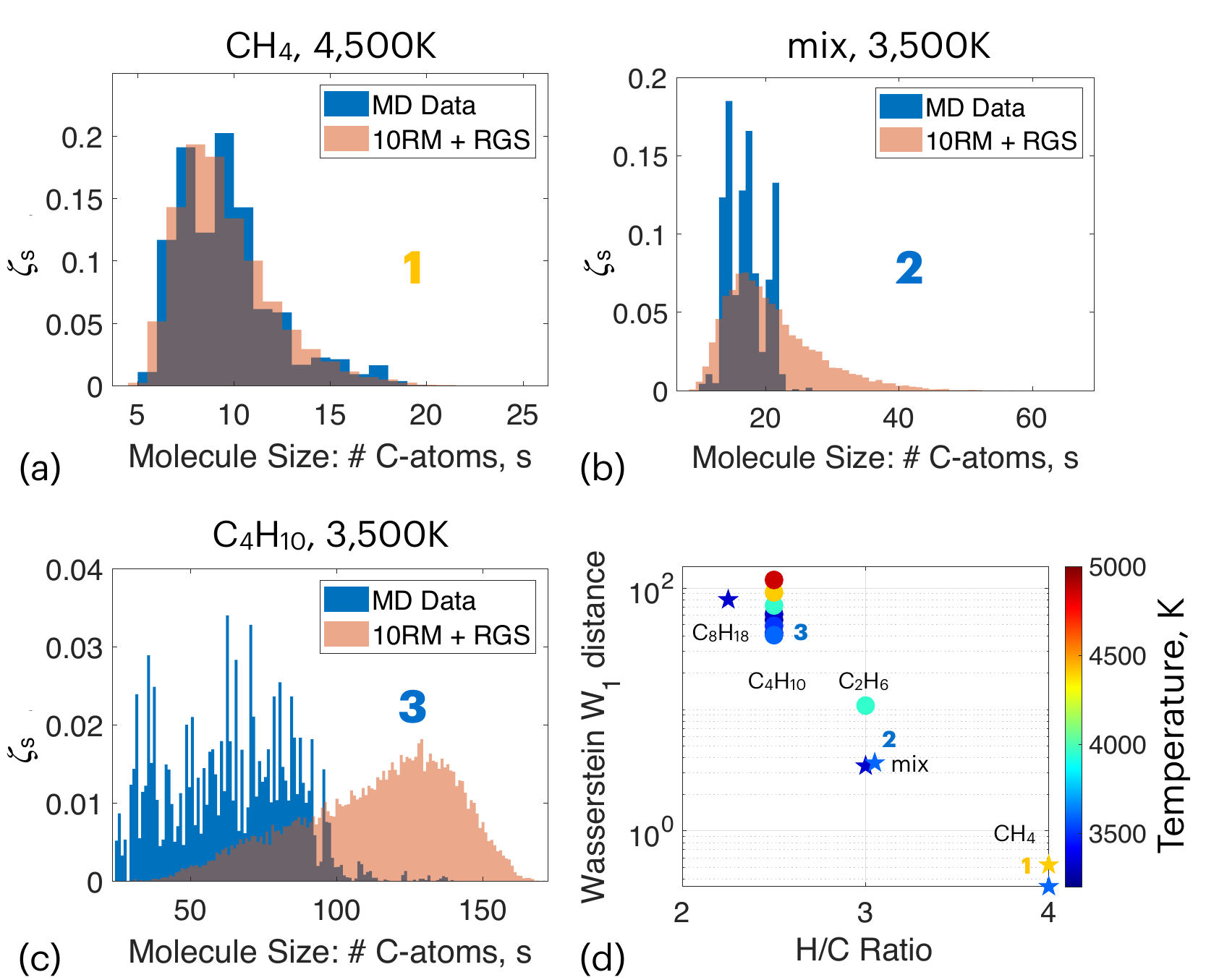}
}   
\caption{(a)-(c) Distributions of the size of the largest molecule extracted from MD data  (blue plots) and obtained using the ten-reaction model and the random graph sampling (10RM+RGS) (red plots). (d) The Wasserstein-1 distance $W_1(P,Q)$ as a function of H/C ratio. The distributions $P$ for the largest molecule size are obtained using the 10RM + RGS model while the distributions $Q$ for the largest molecule size  are extracted from MD simulations. The data are color coded by the temperature of the system.
}
\label{fig:LargeMolHist}
\end{figure*}

Random graph sampling allows us to estimate the probability $\zeta_s$ that the size of the largest molecule is $s$.
We used the 10,000 samples of random graphs to obtain the probability distribution for size of the largest component and compared them to the probability distribution for the largest molecule size extracted from the MD data. This was done for each initial composition and temperature listed in Table \ref{table:MD}. The agreement between these distributions is reasonably good for the cases with high H/C ratio where the giant component does not exist, i.e., for the mix and the initial compositions consisting of only C$_2$H$_6$ or CH$_4$ (see the last five rows in Table \ref{table:giant}). All cases where the giant component was predicted exhibited poor agreement. Examples of good and bad agreements are shown in Figs. \ref{fig:LargeMolHist} (a)--(b) and (c) respectively. 

We would like to comment on the highly noisy distributions for the size of the largest molecule extracted from MD simulations.  First, various MD simulations stayed at equilibrium for different duration ranging from 25\% to 67\% of simulation time. Second, the largest molecule size distribution is naturally very noisy: for example, if a bond is broken in the middle of the largest molecule, its size is divided  roughly by two.

The Wasserstein $W_1$ distance between the size distributions for the largest molecule obtained using random graph sampling and extracted from MD data is plotted in Fig. \ref{fig:LargeMolHist} (d) as a function of H/C ratio. The temperature is shown using color coding. 

Fig. \ref{fig:LargeMolHist} (d) suggests that the prediction for the distribution of the largest molecule size becomes less accurate in terms of the Wasserstein $W_1$ distance as the H/C ratio decreases and the temperature increases.

\section{Discussion}
\label{sec:discussion}
\subsection{Ten-reaction model analysis}

 As shown in Section \ref{sec:res_pred_degr}, the prediction error for the degree distribution by the ten-reaction model increases with temperature. This behavior of the error can be explained by the growth of the number of radicals and double carbon-carbon bonds with temperature. In Fig. \ref{fig:PCbonds}, we have plotted the probability to observe a carbon bonded to a total of $k$ other atoms, hydrogens and carbons, versus temperature. The ten-reaction model relies on Assumption 2 that all carbons are bonded to four other atoms. This assumption becomes less accurate as the temperature increases. At $T=3,200$K, over 96\% of carbons are bonded to four other atoms. At $T=3,600$K, $T=4,000$K and  $T=5,000$K, this percentage drops down to 91\%, 87\%, and 74\%, respectively, due to the increase of the number of double bonds and radicals.

To determine if the errors due to the ten-reaction model are important or not, we compare the small molecule size distributions predicted by the RGT model and the 10RM + RGT model  in Fig. \ref{fig:MSDistr} (right column). Remarkably, these predictions are very close to each other. In particular, the discrepancy between these predictions is very small for the system initially composed of C$_4$H$_{10}$ at 5,000K. Note that the error in the degree distribution predicted by the ten-reaction model for this system was the largest of all systems considered here. This means that the errors observed in Section \ref{sec:res_pred_degr} are low enough and are not a limiting factor for our method in the case of small molecules.

However, in Fig. \ref{fig:LargeMolHist} (d), there is a clear trend that the 10RM + RGS model performs worse for the largest molecule size prediction when the temperature increases. In this model, temperature only affects the 10RM part through the calculation of the $K_{ij}$s. Therefore, for the largest molecule size distribution, it looks like the prediction error from the ten-reaction model is critical and affects the performance of the model. This phenomenon is explained by the fact that errors in the probabilities $p_3$ and $p_4$ for a carbon to have degree three and four, respectively, affect the relative error in size distribution for large molecules the most. To improve the ten-reaction model in future work, assumption 2 could be relaxed and double bonds or radicals could be taken into account.

\subsection{Random graph theory and sampling model analysis}

In Section \ref{sec:results}, we have observed that the RGT and RGS models accurately predict size distributions for small molecules (Fig. \ref{fig:MSDistr}). When no giant component exists, as in the case of systems starting with CH$_4$, C$_2$H$_6$ or the mix, the size of the largest molecules is also well-predicted (see the last five rows of Table \ref{table:giant}, and \ref{fig:LargeMolHist} (a), (b) and (d)). Importantly, the RGT not only predicts the molecule size distribution but also explains it. The distribution is completely determined by three assumptions: $(i)$ the degree distribution, $(ii)$ a large number of carbons, and $(iii)$ equal probability for any pair of carbons to be bonded. These three assumptions imply that small molecules are tree-like with high probability and hence the component size distribution $H_0$ is given by equations \eqref{eq:H1self}--\eqref{eq:H0}.

However, when the random graph theory (specifically, equation  \eqref{eq:crit}) predicts that there is a giant component, the performance of these models worsens, especially for the prediction of the largest molecule (Table \ref{table:giant}, Figs. \ref{fig:LargeMolHist} (c) and (d)). Therefore, \emph{the larger the giant component of the system is, the worse these random graph models perform}. This observation explains why the model consistently performs worse when the H/C ratio decreases. When there are fewer hydrogens per carbon in the system, larger giant components tend to form, resulting in worse predictions by the random graph theory and sampling.

The fact that both the RGT and the RGS model result in a poor prediction for the size of the largest molecule is due to the fact that both of these approaches ignore geometric aspects of the graphs. This means that some large connected components of graphs predicted by the random graph models may be  nonphysical, i.e. impossible to realize in 3D when taking into account the geometric restriction for bond lengths, bond angles, and minimal inter-atomic distance of unbonded atoms. Without these geometric constrains, the molecules predicted by the RGT and RGS models tend to be larger than in the MD simulations, resulting in the poor predictions by the RGT and RGS models. Improvements to the random graph models to take into account these issues will be investigated in future work.

Finally, we would like to make a few remarks on computational cost, the cost of the prediction of molecule size distribution by 10RM+RGT is independent of the size of the system and involves only a solution of a small system of nonlinear equations and function evaluations. Runtimes are about one second. The cost of 10RM+RGS is determined by the number of samples and scales with the number of carbons $N_C$ as the cost of the depth-first search algorithm: $O(N_C)$ \cite{CLRS}. The runtimes for the systems studied in this work are about one minute. It is important to note that 10RM+RGS has an advantage over kinetic Monte Carlo (KMC) hypothetically used with the ten-reaction model that the former would immediately sample random graphs from the invariant distribution while the latter would require some equilibration period. The accuracy of these two techniques would be comparable. Kinetic Monte Carlo combined with a more sophisticated atomic-level reaction model can give accurate predictions even for the distribution of the largest molecule in the system \cite{dufour2021atomic, dufour2022temperature}, but would take more time to run (10s of minutes).

\section{Conclusion}
\label{sec:conclusion}
We have proposed a methodology to predict the molecule size distribution for hydrocarbon pyrolysis suitable for various initial molecular compositions and various temperatures, at least in the range 3200K--5000K. The methodology involves two steps. First, the ten-reaction model analytically gives the degree distribution for the desired initial composition and temperature. Second, given this degree distribution, we predict the molecule size distribution using two random graph theory models. These models can estimate the size distribution of small molecules and of the largest molecule. These two steps are executed within seconds to minutes on a consumer-grade laptop, which is several orders of magnitude faster than what is possible by running MD simulations. Remarkably, the ten-reaction model and the RGT model runtimes do not depend on the size of the system and can easily be computed for any system size. The predictions for small molecule size distributions are remarkably accurate and the predictions for the largest molecule size distributions are good if there are no giant components. Further improvements of the predicting power of the random graph approaches, particularly for the largest molecules, will be performed in future work. Our method could be applied to other systems where the size distribution is of interest. From the first conclusions of our work, our methods can be applied to systems in regimes where no structural effects (cycles, etc) are observed. If molecules become large but no structural effects are observed (only branching), we foresee that our method could be applied successfully as our method assumes that molecules are tree-like. For example, our method could be the case in many polymerization processes.

\section{Supplementary Materials}
The codes developed for this work are available on Github :

\url{https://github.com/mar1akc/RandomGraphTheory4HydrocarbonPyrolysis}.

\subsection{Molecular Dynamics Simulations Details}
\label{sec:MD_details}

Each MD simulation was run in two phases. The goal of phase one was to bring the system to the desired temperature and pressure. The goal of phase two was to let the system reach a chemical equilibrium at the desired temperature and pressure and collect the following data:
\begin{enumerate}
    \item the time-averaged distribution for the numbers of carbon-carbon bonds per carbon atom, or, in other words, the time-averaged degree distribution for the graphs consisting of the union of the carbon skeletons of all molecules present at each time step;
    \item the time-averaged counts of molecules of each size, i.e., the molecule size distribution.
\end{enumerate}

Phase one of each MD simulation started at 111$\,$K and 1013$\,$hPa and ramped up to the desired pressure of 40$\,$GPa and the desired temperature that ranged between 3200$\,$K and 5000$\,$K. The ramping up was spanned over 24$\,$ps with a timestep duration of 0.12$\,$fs. The Nos\'{e}-Hoover thermostat and barostat \cite{NoseHoover1, NoseHoover2, NoseHoover3, NoseHoover4, NoseHoover5} were used with chain length 3 and the damping parameter of 2.4$\,$fs for the temperature and 60$\,$fs for the pressure.

\begin{figure*}[!htb]
  \centering
  \includegraphics[width = 0.6\textwidth]{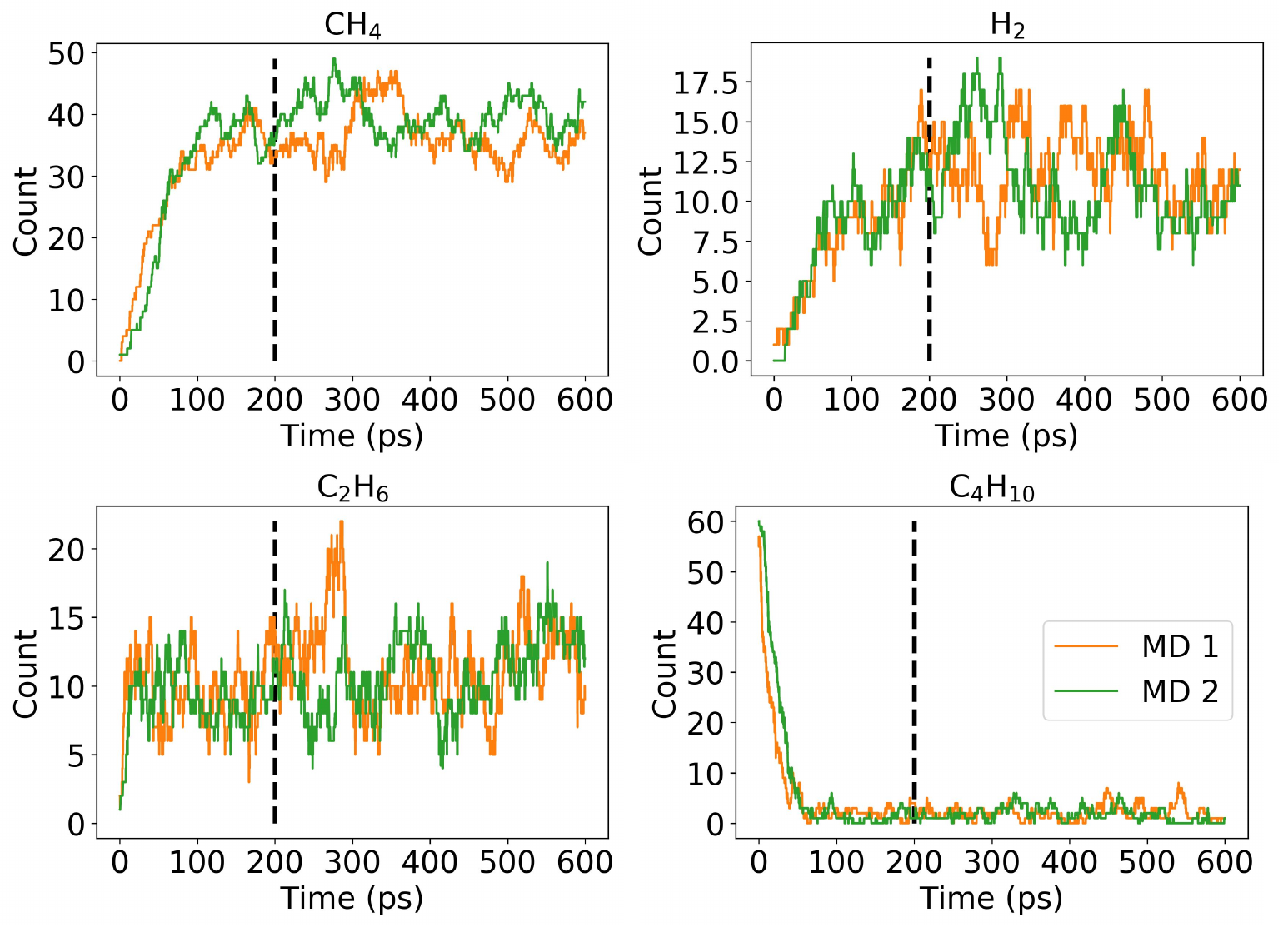}
  \caption{Evolution of the 4 most numerous small molecules (CH$_4$, H$_2$, C$_2$H$_6$, and C$_4$H$_{10}$) for the two MD simulations starting with 64 molecules of C$_4$H$_{10}$ at 3,500$\,$K. Equilibrium is defined when the number of all these 4 molecules are fluctuating around a value for the two MD simulations. Here H$_2$ is the last species to equilibrate and the system is determined to be at equilibrium after 200$\,$ps. Therefore the simulation is 400$\,$ps at equilibrium and this is the value reported in Table \ref{table:MD} for the column ``Sim. time at equilibrium (ps)''.}
  \label{fig:eq_def}
\end{figure*}

Phase two of each simulation was run at constant temperature and pressure. The damping parameters for the Nos\'{e}-Hoover thermostat and barostat were, respectively,  2.4$\,$fs and  14.4$\,$fs. The duration of phase two ranged between 120$\,$ps and 1,000$\,$ps  depending on how long it took for the system to achieve chemical equilibrium. All of the data used in the paper were extracted from phase two, once the equilibrium was reached. Moreover, two MD simulations were run for each set of conditions. Equilibrium was defined as the time where the most numerous small molecules of the system (CH$_4$, H$_2$, C$_2$H$_6$ and C$_4$H$_{10}$) started to fluctuate around a value in the two simulations of the same conditions. Each simulation was left at equilibrium for a long time to have more data and to check that equilibrium was indeed reached. Figure \ref{fig:eq_def} shows an example of how equilibrium was defined and the time at equilibrium for each simulation is given in Table \ref{table:MD}.

\begin{table*}[htb!]
    \centering
    \begin{tabular}{c|c|c|c|c|c|c}
        Init. Comp. &Temperature&$N_C$&$N_H$&H/C&Sim. time (ps) & Sim. time at equilibrium (ps) \\
        \hline
        C$_4$H$_{10}$ & 3200K & 256 & 640 & 2.5 & 600 & 200\\
        C$_4$H$_{10}$ & 3300K & 256 & 640  & 2.5 & 600 & 300 \\
        C$_4$H$_{10}$ & 3400K & 256 & 640  & 2.5 & 600 & 300 \\
        C$_4$H$_{10}$ & 3500K & 256 & 640  & 2.5 & 600 & 400 \\
        C$_4$H$_{10}$ & 3600K & 256 & 640  & 2.5 & 600 & 400\\
        C$_4$H$_{10}$ & 4000K & 256 & 640  & 2.5 & 240 & 150\\
        C$_4$H$_{10}$ & 4500K & 256 & 640  & 2.5 & 120 & 70 \\
        C$_4$H$_{10}$ & 5000K & 256 & 640  & 2.5 & 40 & 30 \\
        C$_2$H$_{6}$ & 3300K & 250 & 750  & 3 & 600 & 300 \\
        C$_2$H$_{6}$ & 4000K & 250 & 750  & 3 & 600 & 150 \\
        C$_8$H$_{18}$ & 3300K & 288 & 648 & 2.25 & 600 & 200 \\
        CH$_{4}$ & 3600K & 160 & 640 & 4 & 1000 & 400 \\
        CH$_{4}$ & 4500K & 160 & 640 & 4 & 240 & 150 \\
        mix & 3500K & 240 & 732 & 3.05 & 600 & 300 \\
    \end{tabular}
    \caption{Conditions for each of the MD simulations. For each of the conditions, two simulations were run. The mix of molecules contained 40 molecules of CH$_4$, 32 molecules of H$_2$, 32 molecules of C$_2$H$_6$, 8 molecules of C$_4$H$_{10}$, 8 molecules of C$_6$H$_{14}$, 4 molecules of C$_8$H$_{18}$, and 2 molecules of C$_{12}$H$_{26}$. ``Sim. time'' is the abbreviation for simulation time. }
    \label{table:MD}
\end{table*}

\subsection{Bond Duration Definition}
\label{sec:bond_duration}

    The bond duration could change on the wide range of temperatures that we are studying; therefore using the value of $\tau = 0.048\,$ps used by Chen et al. \cite{EnzeArticle} may not be appropriate at all temperatures as this value was determined at 3,300$\,$K. In order to study the best value of $\tau$ for different values of temperatures we used the same protocol described by Chen et al. \cite{EnzeArticle}. In short, we compute the relative error between an MD simulation and several simulations of a stochastic simulation algorithm (SSA) fitted on the MD simulation using different values of $\tau$ and take the value of $\tau$ that minimizes this error. The relative error is defined as :
 
    \begin{equation*}
        \label{eq:error_bond_duration}
        {\sf RelErr} = \frac{\sqrt{ \frac{1}{T} \sum_{t=1}^{T} (X_{MD}[t] - \mathbb{E}[{X_{SSA}[t]}])^2}}{\mathbb{E}[\sqrt{ \frac{1}{T} \sum_{t=1}^{T} (X_{i,SSA}[t] - \mathbb{E}[{X_{SSA}[t]}])^2]}},
    \end{equation*}
     where $T$ is the number of timesteps in the MD simulations and the SSA, $X[t]$ is the number of species of interest in the simulation at time $t$. $\mathbb{E}$ is the expected value over 20 SSA. Then, the $l^2$ relative error over the 4 species that are the most common in the system (CH$_4$, H$_2$, C$_2$H$_6$, and C$_4$H$_{10}$):
 
    \begin{equation*}
        \label{eq:l2_error_bond_duration}
        {\sf RelErr}_{l^2} = \sqrt{\sum_{i=1}^4 {\sf RelErr}_i}.
    \end{equation*}
 
\begin{figure*}[!htb]
         \centering
        \includegraphics[width=0.9\textwidth]{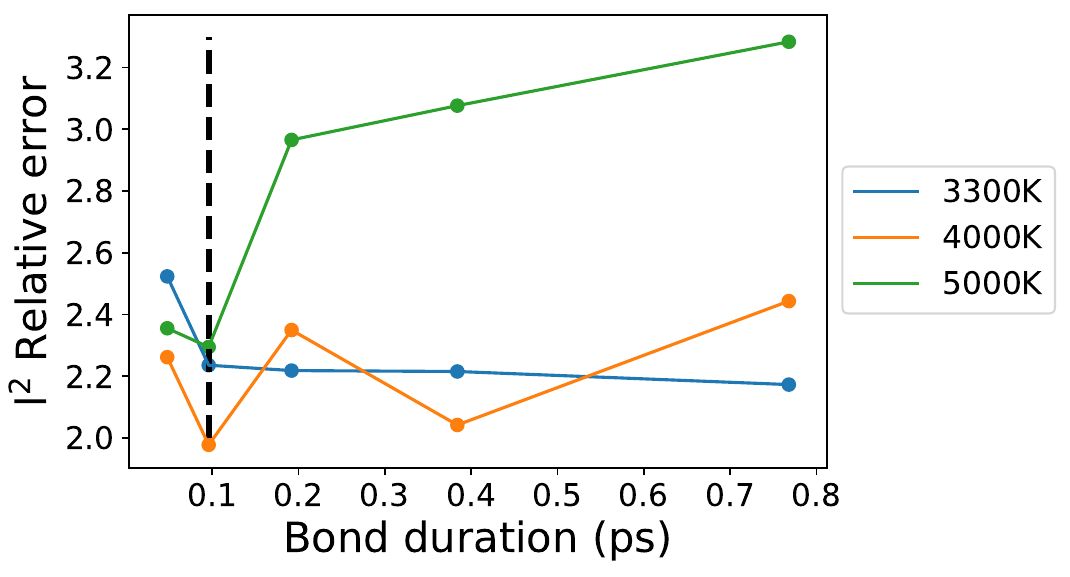}
        \caption{Plot $l^2$ relative errors versus bond duration $\tau$ at different temperatures. The dashed line marks the value $\tau = 0.096$ps which is chosen for this work.
        }
        \label{fig:Bond_duration}
    \end{figure*}

\begin{figure}[!htb]
  \centering
  \includegraphics[width = 0.45\textwidth]{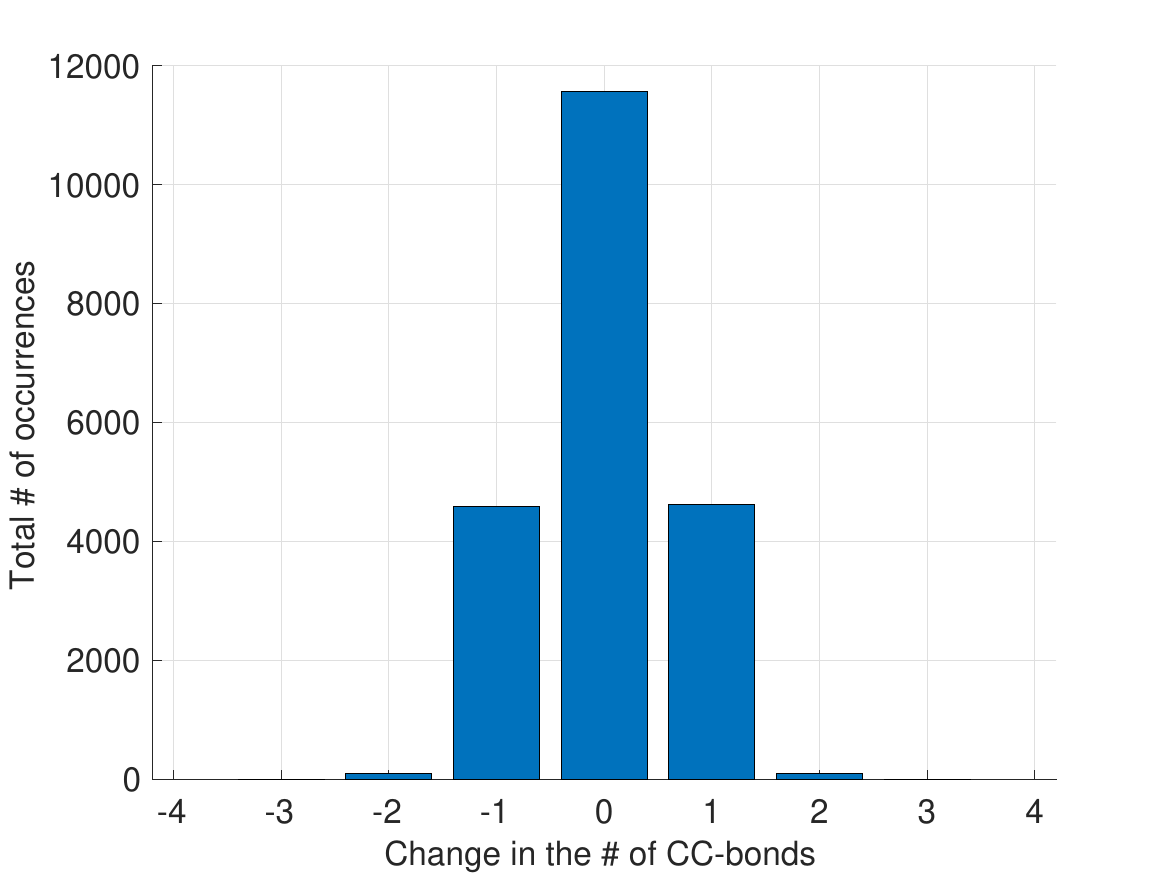}
  \caption{The histogram of changes in the numbers of CC-bonds in chemical reactions observed during MD simulations starting from C$_4$H$_{10}$ at temperature 3300K. }
  \label{fig:CCchangeHist}
\end{figure}

\begin{figure}[!htb]
  \centering
  \includegraphics[width = 0.45\textwidth]{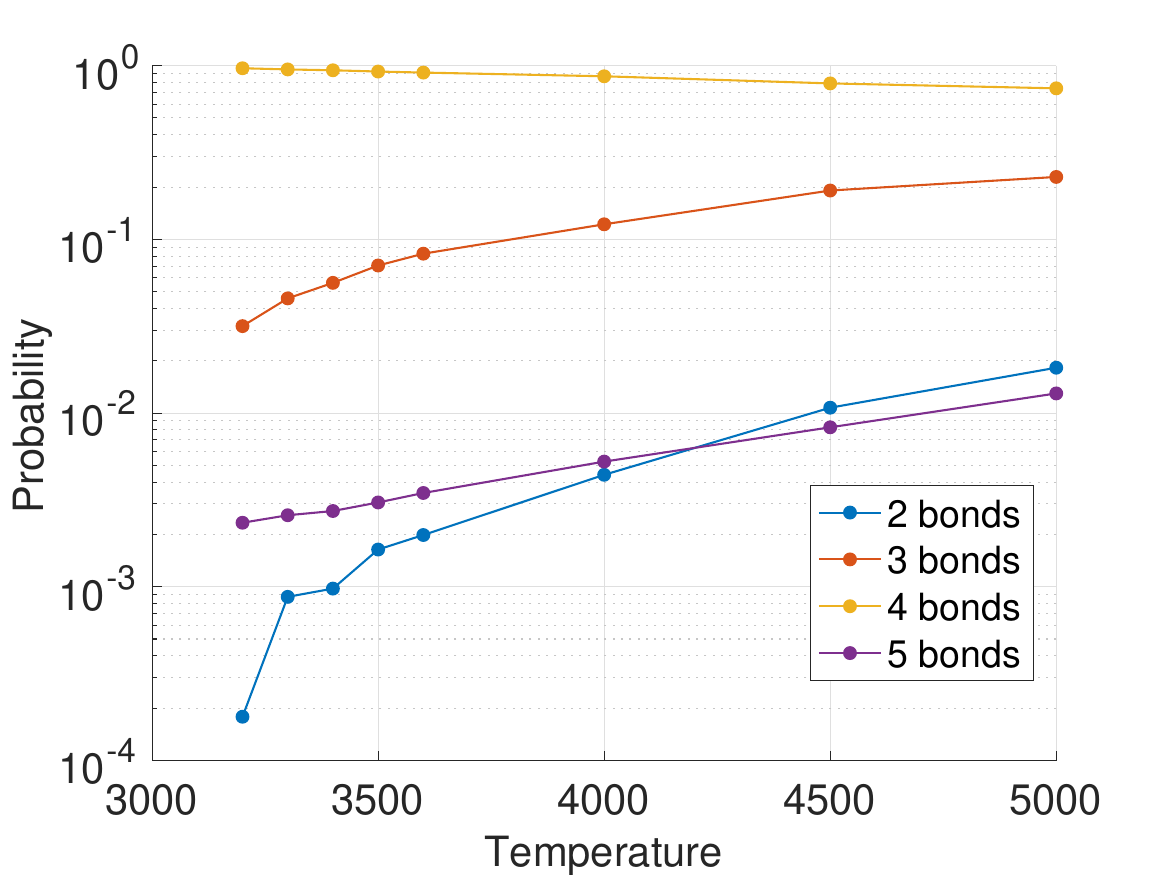}
  \caption{The probability to observe a carbon atom bonded to a total of 2,3,4, or 5 other atoms.}
  \label{fig:PCbonds}
\end{figure}

     We perform this analysis on 3 simulations starting with C$_4$H$_{10}$ at different temperatures (3,300, 4,000, and 5,000$\,$K), and the results are shown in Fig. \ref{fig:Bond_duration}. We can see that $\tau = 0.096$ps is the best value for 4,000 and 5,000$\,$K and that there is only a small change above 3,300$\,$K in relative error above $\tau = 0.096$ps, so for consistency, we chose this value of bond duration for all temperatures.
 
\pagebreak
\subsection{Ten-reaction model details}
\label{sec:10RM_details}
\subsubsection{Ten-reaction model construction}

\pagebreak

\subsubsection{Calibrating the ten-reaction model}
The numbers $K_{ij}$ denote the equilibrium constants of the reaction defined by:
\begin{equation}
   K_{ij} = \frac{[C_{i + 1} - C_{j + 1}]_{\rm eq}[H-H]_{\rm eq}}{[C_i-H]_{\rm eq}[C_j-H]_{\rm eq}},
   \label{eq:equilibrium_constant}
\end{equation}
where [X]$_{\rm eq}$ is the concentration of the species X at equilibrium, i.e., the ratio of the mean number of species X in the simulation to the volume of the simulation cell. For convenience, the subscript `eq' will be dropped from now on since only the equilibrium conditions are considered here.
Since the volume of the simulation cell is the same for all species, it can be canceled out in Eq. \eqref{eq:equilibrium_constant} resulting in

\begin{equation}
   K_{ij} = \frac{N_{C_{i + 1} - C_{j + 1}}N_{H-H}}{N_{C_i-H}N_{C_j-H}}.
   \label{eq:equilibrium_constant_simplified}
\end{equation}
The number $N_{C_{i + 1} - C_{j + 1}}$ is the time-averaged number of bonds between carbons of degrees $i+1$ and $j+1$ at equilibrium. The numbers $N_{H-H}$ and $N_{C_i-H}$ are defined similarly. While these numbers can be extracted from MD simulations directly, it is more convenient to extract  $N_{C_i}$, the time-averaged counts of carbons with degrees $i$ at equilibrium,  $i=0, 1, 2, 3, 4$, and express  $N_{C_{i + 1} - C_{j + 1}}$ and  $N_{C_i-H}$ via them. In addition, expressing $N_{C_{i + 1} - C_{j + 1}}$ and  $N_{C_i-H}$ in function of the $N_{C_i}$s will be necessary to reduce the number of variables in Section \ref{sec:computeDD10RM}.

\label{sec:calib}
Assumption 2 implies that a carbon with degree $i$ forms $4-i$ bonds with hydrogens. This yields a total of 
\begin{equation}
   N_{C_i-H} = (4 - i)N_{C_i}
   \label{eq:NC}
\end{equation}
$(C_i-H)$-bonds in the system. Since all carbons have equal probabilities to be found at the end of a randomly chosen CC-bond, the probability $\rho_i$ that one of the ends of a random CC-bond has degree $i$ is:
\begin{equation}
   \rho_i = \frac{i N_{C_i}}{N_{C_1} + 2 N_{C_2} + 3N_{C_3} + 4N_{C_4}}.
   \label{eq:probability_stub}
\end{equation}
Therefore, the numbers $N_{C_{i} - C_{j}}$ are equal to:
\begin{equation}
   N_{C_{i} - C_{j}} = \begin{cases}2\rho_i\rho_jN_{C-C},& i\neq j,\\
   \rho_i^2N_{C-C},& i=j,
   \end{cases}
   \label{eq:NCiCj}
\end{equation}
where $N_{C-C}$ is the total number of CC-bonds at equilibrium:
\begin{equation}
   N_{C-C} = \frac{N_{C_1} + 2 N_{C_2} + 3N_{C_3} + 4N_{C_4}}{2}.
   \label{eq:NCC}
\end{equation}

Equations \eqref{eq:equilibrium_constant_simplified}--\eqref{eq:NCC} enable us to rewrite the equilibrium constant as:
\begin{equation}
   K_{ij} = \begin{cases}\frac{2(i+1)(j+1)N_{C_{i+1}}N_{C_{j+1}}N_{H-H}}{4N_{C-C}(4-i)N_{C_i}(4-j)N_{C_j}}, & i\neq j\\
    \frac{(i+1)^2N_{C_{i+1}}^2N_{H-H}}{4N_{C-C}(4-i)^2N_{C_i}^2},& i=j.
    \end{cases}
   \label{eq:equilibrium_constant_final_ij}
\end{equation}
Therefore, the reaction rate constants $K_{ij}$ can be evaluated using only the time-averaged numbers $N_{C_i}$, $N_{C-C}$ and $N_{H-H}$ extracted from MD data.

We make the common assumption that the $K_{ij}$s follow the Arrhenius law
\begin{equation}
   K_{ij} =A_{ij}e^{\frac{-B_{ij}}{T}}, 
   \label{eq:arrhenius}
\end{equation}
where $T$ is the temperature of the simulation, $A_{ij}$ and $B_{ij}$ are the Arrhenius parameters. To determine these parameters, we extracted the time averages of $N_{C_i}$, $i = 0, 1, 2, 3,$ or $4$, $N_{C-C}$ and $N_{H-H}$ from MD simulations starting with 64 molecules C$_4$H$_{10}$ at temperatures ranging from 3200K to 5000K, calculated the constants $K_{ij}$, and fitted them to the Arrhenius law.  Importantly, the constants $A_{ij}$ and $B_{ij}$ are considered independent of the chemical composition of the system. Therefore, we expect that the found constants $A_{ij}$ and $B_{ij}$ will be suitable for other initial compositions as well. The constants $A_{ij}$ and $B_{ij}$ for the ten reactions are reported in Table \ref{table:arrhenius}. 

\begin{table}[!htb]
    \centering
    \begin{tabular}{c|c|c}
        Eq. Const. & A$_{ij}$ & B$_{ij}$\\
        \hline
        K$_{00}$ & 1.26 & 8742\\
        K$_{01}$ & 1.47 & 8294\\
        K$_{02}$ & 1.10 & 7229\\
        K$_{03}$ & 0.59 & 5146\\
        K$_{11}$ & 0.93 & 7847\\
        K$_{12}$ & 0.94 & 6782\\
        K$_{13}$ & 0.50 & 4698\\
        K$_{22}$ & 0.52 & 5717\\
        K$_{23}$ & 0.38 & 3633\\
        K$_{33}$ & 0.15 & 1549\\
    \end{tabular}
    \caption{Arrhenius parameters, as defined in Eq. \ref{eq:arrhenius}, for the equilibrium constants $K_{ij}$ of the ten-reaction model. ``Eq. Const.'' is the abbreviation for ``Equilibrium Constant'', A$_{ij}$ is nondimensional, and B$_{ij}$ is in K$^{-1}$.}
    \label{table:arrhenius}
\end{table}

\textbf{\begin{figure*}[!htb]
  \centering
  \includegraphics[width=0.9\textwidth]{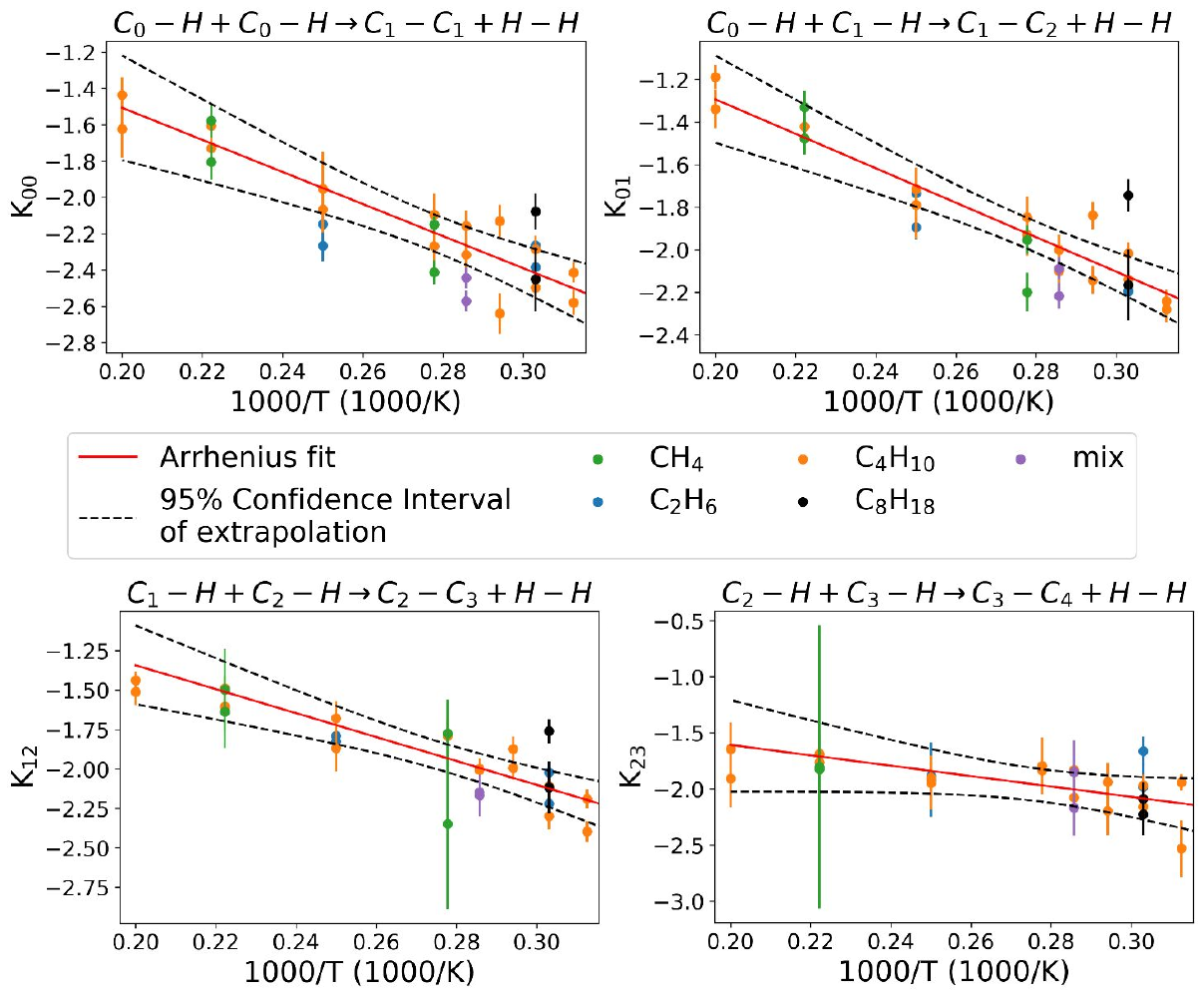}
  \caption{Plots of the equilibrium constants (log scale) versus the inverse of the temperature for the 4 reactions that are used to obtain the degree distributions. The red line are fitted to the C$_4$H$_{10}$ data (orange). The Arrhenius law for the equilibrium constants is a good approximation on the range 3200K-5000K; especially, even when changing the starting molecules the equilibrium constants still follow the Arrhenius law. The green dot corresponding to the MD simulation initialized with 160 molecules CH$_4$ at 3,300$\,$K is missing in the plot for $K_{23}$ as this simulation had not yielded any carbon with degree 4.}
  \label{fig:K_vs_T}
\end{figure*}}

\subsubsection{Computing the degree distribution}
\label{sec:computeDD10RM}
In this section, we explain how to obtain $N_{C_i}$, $0\le i\le 4$, the expected counts of carbon atoms  with degree $i$,  for a given initial composition and a given temperature. The degree distribution is then readily computed:
\begin{equation}
    \label{eq:pi}
    p_i=\frac{N_{C_i}}{N_C},\quad i=0,1,2,3,4,
\end{equation}
where $N_C$ is the total number of carbons in the system.

While there are ten reaction constants $K_{ij}$, only four of them are functionally independent. Indeed, all of them are functions of the five ratios
$$
\frac{N_{H-H}}{N_{C-C}}\quad{\rm and}\quad \frac{N_{C_{i+1}}}{N_{C_i}},~i=0,1,2,3.
$$
Therefore, we can express $N_{C_1}$, $N_{C_2}$, $N_{C_3}$, and $N_{C_4}$ via $N_{C_0}$, $N_{H-H}$, $N_{C-C}$ and four of the $K_{ij}$. We have chosen to use $K_{00}$, $K_{01}$, $K_{12}$, and $K_{23}$:
\begin{align}
   N_{C_1} &=8\sqrt{K_{00}}N_{C_0}\sqrt{\frac{N_{C-C}}{N_{H-H}}}, 
   \label{eq:NC1}\\
   N_{C_2} &=12K_{01}N_{C_0}\frac{N_{C-C}}{N_{H-H}}, 
   \label{eq:NC2}\\
   N_{C_3} &=16\sqrt{K_{00}}K_{12}N_{C_0}\left(\frac{N_{C-C}}{N_{H-H}}\right)^{3/2}, 
   \label{eq:NC3}\\
   N_{C_4} &=4K_{01}K_{23}N_{C_0}\left(\frac{N_{C-C}}{N_{H-H}}\right)^{2}.
   \label{eq:NC4}
\end{align}

At this point, three variables are unknown: $N_{C_0}$, $N_{H-H}$, and $N_{C-C}$; therefore, three additional equations are needed. Observing that the total numbers of carbons, $N_C$, and hydrogens, $N_H$, remain constant during any MD simulation, we obtain:
\begin{align}
   N_{C} &=N_{C_0} + N_{C_1} + N_{C_2} + N_{C_3} + N_{C_4},
   \label{eq:number_of_C}\\
   N_{H} &= 4N_{C_0} + 3N_{C_1} + 2N_{C_2} + N_{C_3} + 2N_{H-H}.
   \label{eq:number_of_H}
\end{align}
The relationship between the counts of CC-bonds and HH-bonds at any moment of time and their initial counts follows from the fact that, in our model, the creation (removal) of a CC-bond is always accompanied with the creation (removal) of an HH-bond:
\begin{equation}
   N_{H-H} - N_{H-H}^{init} = N_{C-C} - N_{C-C}^{init}.
   \label{eq:NHH_and_NCC}
\end{equation}
Therefore, as we know the temperature $T$, the Arrhenius constants $A_{ij}$ and $B_{ij}$, we can calculate $K_{00}$, $K_{01}$, $K_{12}$, and $K_{23}$. And with the initial conditions $N_C$, $N_H$, $N_{H-H}^{init}$, $N_{C-C}^{init}$, we can solve the system of nonlinear equations \eqref{eq:NC1}--\eqref{eq:NHH_and_NCC} and obtain $N_{C_0}$, $N_{C_1}$, $N_{C_2}$, $N_{C_3}$, $N_{C_4}$.  Then the desired degree distribution is found using \eqref{eq:pi}. 

\subsubsection{Error calculation for degree distribution}

 The uncertainties of the prediction of the degree distribution result from the uncertainties in the values of $N_{C_i}$, $0\le i\le 4$, $N_{H-H}$, and $N_{C-C}$ extracted from the MD data. 
 In this section, we detail uncertainty quantification for the degree distribution. This is done in two stages: we estimate the uncertainties in $K_{ij}$s first and then use them to find the uncertainty in the degree distribution. 
 
We adopt a reasonable assumption that $N_{C_i}$, $N_{H-H}$, $N_{C-C}$, and $K_{ij}$ obey the log-normal distributions. The mean values and the standard deviations of the logarithms of $N_{C_i}$, $0\le i\le 4$, $N_{H-H}$, and $N_{C-C}$ are computed over the last 10\% of each MD simulation.
We collect $M = 1000$ samples of each quantity $N_{C_i}^{T_{train}}$, $N_{H-H}^{T_{train}}$, and $N_{C-C}^{T_{train}}$ from the last 10\% of the MD simulation at temperature $T_{train}$  and fit them to log-normal distributions. Next, we calculate the corresponding $M$ values of $K_{ij}^{T_{train}}$ at each temperature $T_{train}$ using Eq. (\ref{eq:equilibrium_constant_final_ij}) and fit them to log-normal distributions. 
 The mean values and the standard deviations of $K_{ij}^{T_{train}}$ are shown with the orange points and the orange error bars in Fig. \ref{fig:K_vs_T}.
 Then we fit the Arrhenius law Eq. \ref{eq:arrhenius} to the collection of the mean values of K$_{ij}^{T_{train}}$ (the red line in Fig. \ref{fig:K_vs_T}) and determine the parameters $A_{ij}$ and $B_{ij}$. 
 
The value of $K_{ij}^{T_{test}}$ at a new temperature value $T_{test}$ determined from this fit involves an uncertainty.  In order to quantify it, the bootstrap method  \cite{efron1994introduction} is used.
For each pair of values of $i$ and $j$, $M=1000$ values of the $K_{ij}^{T_{train}}$ at all training temperatures are sampled from their respective log-normal distributions and fitted to the Arrhenius law (Eq. (\ref{eq:arrhenius})). Then the confidence distribution of extrapolation
of each of these plots is computed using the standard statistical calculations \cite{buteikis2019practical}. The mean of the means and the mean of the standard deviations of the $M$ confidence distributions are computed, and the confidence distribution of extrapolation is obtained. The 95\% confidence intervals of the confidence distribution of extrapolation are shown in black in Fig. \ref{fig:K_vs_T}.
 
 Finally, the uncertainty of the degree distribution is computed using these confidence distributions of extrapolation using the bootstrap. At the specific temperature $T_{test}$ that is being studied, $M = 1000$ values of the four $K_{ij}$s Eqs.  (\ref{eq:NC1})-(\ref{eq:NC4}) are sampled from the confidence distribution of extrapolation. The system of equations (Eqs. (\ref{eq:NC1})-(\ref{eq:NHH_and_NCC})) is then solved $M$ times yielding $M$ values for the degree counts $N_{C_i}$. The standard deviation for each  $N_{C_i}$ gives the standard error of $N_{C_i}$ shown with the black error bars in Fig. \ref{fig:Degrees_pred}.

\subsection{Generating molecule size distribution from degree distribution (a detailed version)}
\label{sec:RGT_details}
\subsubsection{The configuration model for carbon skeletons }
Since the valency of carbon is 4, there are only 5 nonzero entries in the \emph{degree distribution}: $p_0$, $p_1$, $p_2$, $p_3$,  and  $p_4$, where $p_k$ is the probability that a randomly picked C-atom has $k$ CC-bonds \footnotemark[1].
 \footnotetext[1]{While carbon atoms with 5 carbon-carbon bonds are occasionally observed in MD simulations, they exist for very short times. Hence, we neglect them.} 
Given the degree distribution and assuming that the number of vertices (C-atoms) is very large, one can use the elegant approach of Newman, Strogatz, and Watts \cite{NSW} based on generating functions to obtain the probabilities $P_s$, $s\in\mathbb{N}$, that a randomly chosen C-atom belongs to a molecule containing $s$ C-atoms. MD data allow us to estimate the probabilities 
 $\pi_s$ that a randomly picked molecule has $s$ C-atoms. The distribution $\{\pi_s\}$ is readily obtained from the distribution $\{P_s\}$:
\begin{equation}
    \label{eq:pis_SM} \pi_s = \frac{P_s/s}{\sum_{j\in\mathbb{N}} P_j/j},\quad s\in\mathbb{N}.
\end{equation}

Now we will walk the reader through the steps of the generating function approach \cite{NSW} and apply it to hydrocarbons. For any discrete probability mass function (PMF)  $\{a_k\}$, the generating function is defined by $G(x):=\sum_{k} a_kx^k$. Generating functions possess a number of useful properties \cite{Wilf}.  Since any PMF sums up to 1, we have $G(1) = 1$.
The definition of the mean implies that $\langle k\rangle = G'(1)$. Given a generating function, one can restore the probability distribution:
\begin{equation}
    \label{eq:restore}
    a_k = \frac{1}{k!}\left.\frac{d^k G}{d x^k} \right|_{x=0} = \frac{1}{2\pi i}\oint\frac{G(z)dz}{z^{k+1}}.
\end{equation}
The last expression, Cauchy's integral formula, gives a practical tool for obtaining the distribution in the case where the analytical expression for $G$ is complicated or does not exist. The integral is taken along an arbitrary circle around the origin in the complex plane that bounds a disc within which $G$ is holomorphic.

The generating function for the degree distribution is denoted by $G_0(x)$. In our case, $G_0(x)$ is a polynomial of degree 4:
\begin{equation}
    \label{eq:G0}
    G_0(x)=p_0+p_1x+p_2x^2+p_3x^3+p_4x^4.
\end{equation}
  Besides the degree distribution $p_k$, we also need the so-called \emph{excess degree distribution} $\{q_k\}$ defined as follows. We randomly pick an edge and then randomly pick one of its endpoints $v$.  The probability that $v$
has $k$ other edges incident to it is 
\begin{equation}
\label{eq:qk}
q_k = \frac{(k+1)p_{k+1}}{G_0'(1)}.
\end{equation}
Note that $q_k = \rho_{k+1}$ where $\rho$ is defined in \eqref{eq:probability_stub}.
The distribution $\{q_k\}$ is generated by the function $G_1(x) = G'_0(x)/G'_0(1)$.
In our case, the function $G_1(x)$ is a cubic polynomial $G_1(x) = q_0+q_1x+q_2x^2+q_3x^3$.

Each random graph with a large number of vertices $n$ and a prescribed degree distribution $\{p_k\}$ is in one of two phases \cite{MR1995,NSW}:
\begin{equation}
\label{eq:crit_SM}
    \sum k(k-2)p_k\begin{cases} <0,&\text{subcritical phase}\\> 0,&\text{supercritical phase}\end{cases}.
\end{equation}
In the subcritical regime, all connected components of the graph are small, while in the supercritical regime, there is a unique \emph{giant component} containing $O(n)$ vertices, and the rest of connected components (if any) are small. The probability distribution $\{P_s\}$ for the sizes of small components is generated by the function $H_0(x)$. 
This function can be found from the generating function $H_1(x)$ for the probability distribution for the component sizes hanging out from one of the ends of a randomly picked edge. The function $H_1(x)$ satisfies the self-consistency relationship \cite{NSW} 
\begin{equation}
    \label{eq:H1self_SM}
    H_1(x) = xG_1(H_1(x)).
\end{equation}
Equation \eqref{eq:H1self_SM} follows from the assumption that all small components are tree-like (which holds with probability tending to 1 as $n\rightarrow\infty$) and the power property of generating functions: if $G(x)$ generates the distribution for some quantity associated with an object, then $[G(x)]^m$ generates the distribution for the sum of these quantities of $m$ independent samples of the object (see Fig. \ref{fig:self} in the main text). 
In our case, equation \eqref{eq:H1self_SM} is cubic in $H_1$:\begin{equation}
    \label{eq:H1}
    H_1 = xq_0 + xq_1H_1 + xq_2H_1^2 + xq_3H_1^3.
\end{equation}

Equation \eqref{eq:H1} has three roots. One of them tends to zero as $x\rightarrow0$ while the other two blow up as $x\rightarrow0$. Note that $H_1(0)$ is the probability that the connected component attached to an arbitrary end of a randomly picked edge contains zero vertices which is impossible. Hence $H_1(0)$ must be zero. Therefore, we select the root with the smallest absolute value.
Having solved \eqref{eq:H1} for $H_1$, we find $H_0(x)$ from the formula \cite{NSW} (see Fig. \ref{fig:self} (bottom) in the main text)

\begin{equation}
    \label{eq:H0_SM}
    H_0(x)=xG_0(H_1(x))
\end{equation}
and restore the distribution $\{P_s\}$ using \eqref{eq:restore}. Finally, we obtain the desired distribution $\pi_s$ for the size of a randomly picked connected component using \eqref{eq:pis_SM}. 

The criterion \eqref{eq:crit_SM} for the existence of a giant component can be simplified for the case of hydrocarbons since the highest degree is $4$:

\begin{equation}
    \label{eq:hydrocrit}
    \begin{cases}-p_1+3p_3 + 8p_4 < 0, \text{subcritical phase}\\-p_1+3p_3 + 8p_4 > 0, \text{supercritical phase.}
    \end{cases}
\end{equation}
When the system is in the supercritical phase, we can estimate the fraction $S$  of vertices in the giant component using the relationship \cite{NSW}:
\begin{equation}
    \label{eq:largestcompsize}
    H_0(1) = G_0(H_1(1)) = 1-S.
\end{equation}
The existence of the giant component in our case suggests the presence of a large molecule containing substantial fraction of carbons at all times.
Note that in the supercritical phase we have $H_1(1) = u<1$, where $u$ is the smallest positive solution to $u=G_1(u)$.
Hence, the expected number of carbon atoms in the largest hydrocarbon molecule is $N_C(1-H_0(1))$.  

\subsubsection{Summary}

The molecule size distribution $\pi_s$, $s\in\mathbb{N}$, is obtained from the degree distribution $p_k$, $k=0,1,2,3,4$ as follows.
\begin{enumerate}
    \item First, the excess degree distribution $q_k$, $k = 0,1,2,3$, is obtained using \eqref{eq:qk} and the generating function $G_1(x)$ is formed.
    \item Then the generating function $H_1(x)$ is computed by solving the cubic equation \eqref{eq:H1}.
    \item Then the generating function $H_0(x)$ for the sizes of non-giant components is defined by \eqref{eq:H0_SM}.
    \item Next, the coefficients $P_s$ for the power series expansion of $H_0(x)=\sum_{s=1}^{\infty}P_sx^s$ are computed using Cauchy's integral formula \eqref{eq:restore}:
    \begin{equation}
        \label{eq:Cauchy}
        P_s = \frac{1}{2\pi i}\oint_{|z|=1}\frac{H_0(z)}{z^{s+1}}.
    \end{equation}
    It is important to remember that $P_s$ is the probability for a randomly picked carbon atom to belong to a molecule with $s$ carbon atoms, i.e., molecule of size $s$ in our terminology. The distribution $\{P_s\}$ sums up to $1-S$ where $S$ is the expected fraction of atoms belonging to the giant component. 
    \item Finally, the probabilities $\pi_s$ for a randomly picked carbon-containing molecule to be of size $s$ are computed from the distribution $\{P_s\}$ using \eqref{eq:pis_SM}.
\end{enumerate}

 \subsubsection{Error estimation for the molecule size distribution}
The degree distribution $\mathbf{p}:=[p_0,p_1,p_2,p_3,p_4]^\top$ for the carbon skeletons is not known precisely but can be either extracted from MD simulations or predicted as described in Section III A. Both of these approaches involve errors:
\begin{equation}
    \label{pek}
    p_k\pm e_k,\quad 0\le k\le 4.
\end{equation}
If the probability $p_k$ is estimated from MD data, the error $e_k$ is taken to be two standard deviations for $p_k$.

We estimate the error $\epsilon_s$ for $\pi_s$, $s=1,2,\ldots$ by setting
\begin{equation}
    \label{eq:error_pis}
    \epsilon_s = \sum_{k=0}^4\left|\frac{\partial \pi_s}{\partial p_k}\right|e_k.
\end{equation}
The partial derivatives in \eqref{eq:error_pis} are found as follows. Since $\{p_k\}_{k=0}^4$ is a probability mass function, it must sum up to 1. Hence, if we change one of $p_k$, the others should also change in such a manner that the sum of $p_k$ remains 1.  To account for this aspect, we find an orthonormal basis $B:=[\mathbf{b}_1,\mathbf{b}_2,\mathbf{b}_3,\mathbf{b}_4]$ in the hyperplane $p_0+p_1+p_2+p_3+p_4=1$.
Next, we approximate the partial derivatives of $\pi_s$ with respect to each $\mathbf{b}_j$, $j=1,2,3,4$, by finite differences:
\begin{equation}
    \label{eq:dpis/dbj}
    \frac{\partial \pi_s}{\partial \mathbf{b}_j}\approx \frac{\pi_s(\mathbf{p}+\delta \mathbf{b}_j) -\pi_s(\mathbf{p}-\delta \mathbf{b}_j)}{2\delta}. 
\end{equation}
These partial derivatives comprise a four-component vector $\nabla_{\mathbf{b}} \pi_s$.
Then the desired derivatives $\pi_s$ with respect to $p_k$, $0\le k\le 4$, are
\begin{equation}
    \label{eq:dpis/dpk}
    \nabla_\mathbf{p} \pi_s:=\left[\frac{\partial \pi_s}{\partial p_0},\frac{\partial \pi_s}{\partial p_1},\frac{\partial \pi_s}{\partial p_2},\frac{\partial \pi_s}{\partial p_3},\frac{\partial \pi_s}{\partial p_4}\right]^\top = B\nabla_{\mathbf{b}} \pi_s.
\end{equation}
It is easy-to-check that the components of $\nabla_\mathbf{p} \pi_s$ sum up to zero.


\subsection{Additional plots for the size distribution for small molecules} 
\label{sec:small_mol_fig} 
\begin{figure*}[htb]
  \centerline{
 (a) \includegraphics[width=0.9\textwidth]{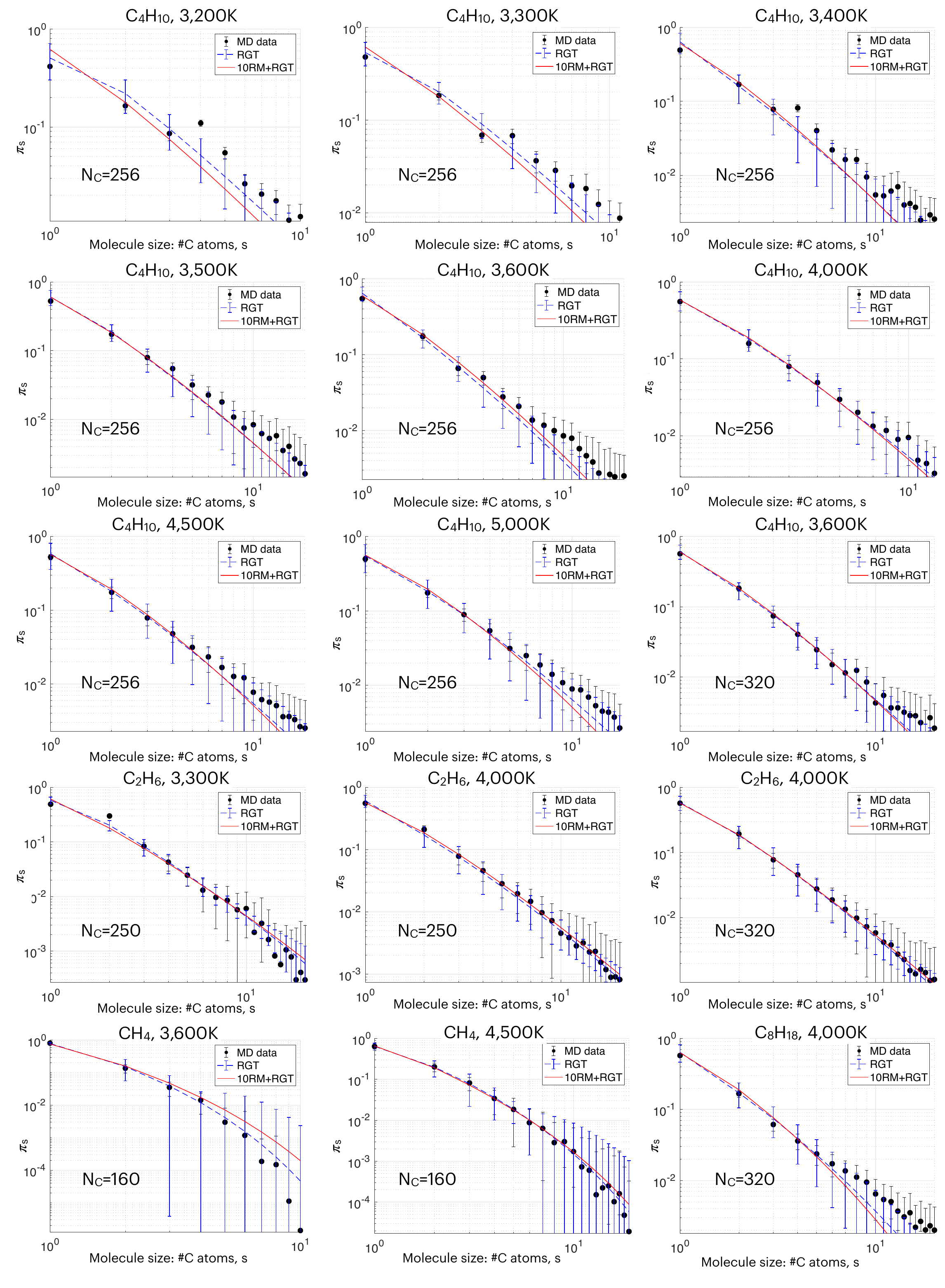}
 }
  \caption{Molecule size distributions for molecules containing up to 20 carbons using random graph theory and the ten-reaction model. The results are shown for various initial compositions and temperatures. This figure complements Fig. 4 in the main text.
  } 
  \label{fig:small_mol_fig}
\end{figure*}

A sample of four predicted size distributions for small molecules by RGT and 10RM+RGT is shown in Fig. 4 in the main text. Fig. \ref{fig:small_mol_fig} displays  predicted size distributions for small molecules for all other systems studied in this work.

\pagebreak

\subsection{System size effect}
\label{sec:size_effect}

\begin{figure*}[!htb]
  \centerline{
 (a) \includegraphics[width=\textwidth]{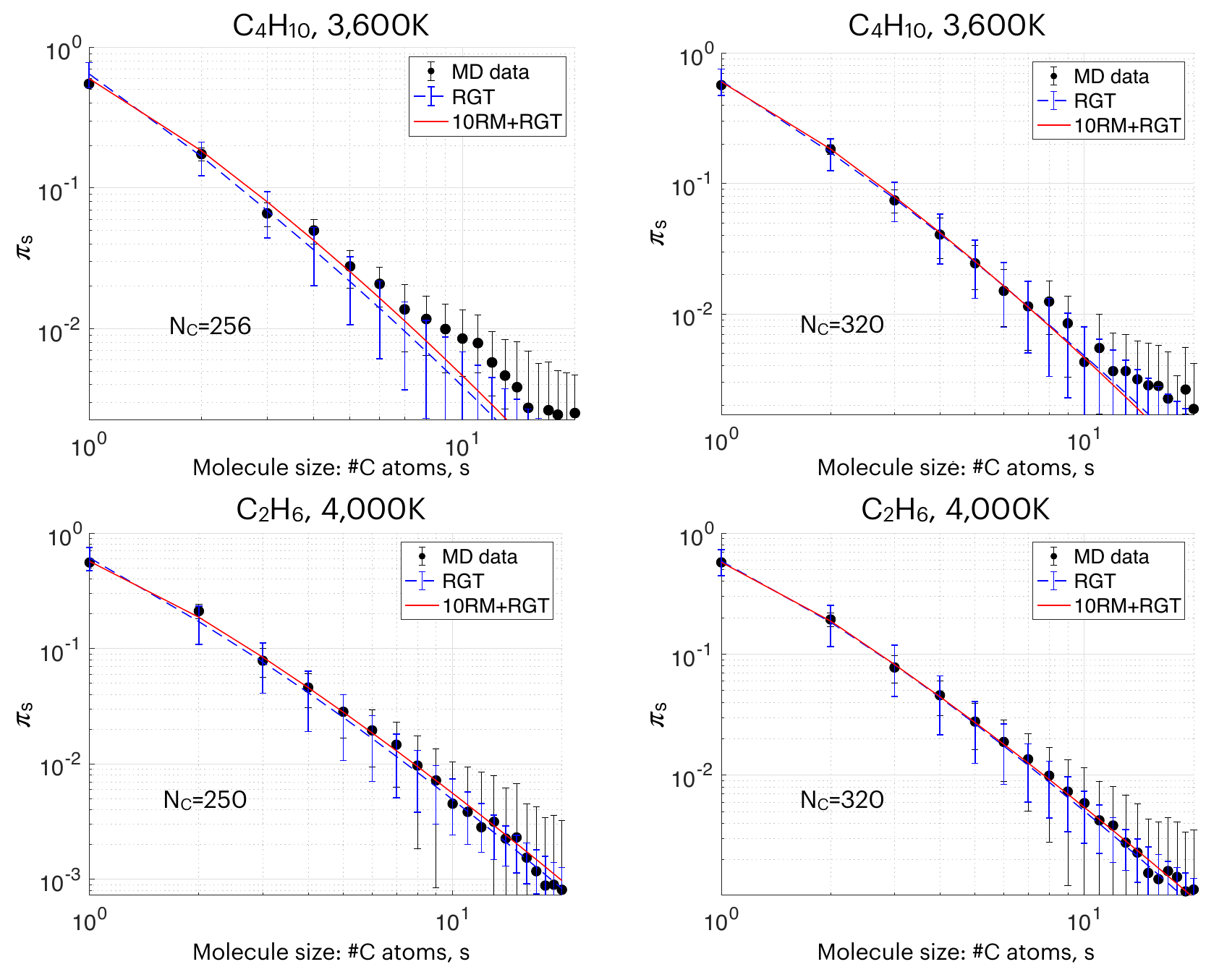}
 }
  \caption{Molecule size distributions for molecules containing up to 20 carbons using random graph theory and the ten-reaction model.  This figure illustrates the accuracy of the predictions for the molecule size distribution on the size of the system.
  } 
  \label{fig:size_effect}
\end{figure*}

The formulas for the small component size distributions are derived under the assumption that the number of vertices of the graph is very large. Therefore, we expect that the increase of the number of carbons in the system will improve the accuracy of the prediction for the size distribution for small molecules. Fig. \ref{fig:size_effect} demonstrates this effect. Its top row displays the results for two systems initialized with C$_4$H$_{10}$ molecules at temperature 3,600K containing, respectively, a total of 256 and 320 carbons. The results for the other two systems initialized with C$_2$H$_6$ at temperature 4,000K are shown in the bottom row. The predictions for the larger systems by RGT and 10RM+RGT are more accurate. The corresponding values of the Wasserstein $W_1$ distances reported in Table \ref{table:size_effect} quantify this effect.
\begin{table*}[!htb]
    \centering
    \begin{tabular}{c|c|c|c|c}
        Init. Comp. & Temperature & \# carbons, $N_C$ & $W_1({\rm MD},{\rm RGT})$ &  $W_1({\rm MD},{\rm 10RM+RGT})$ \\
         \hline
         C$_4$H$_{10}$ & 3,600K & 256 & 0.58 & 0.43 \\
         C$_4$H$_{10}$ & 3,600K & 320 & 0.23 & 0.23 \\
         \hline
         C$_2$H$_{6}$ & 4,000K & 250 & 0.12 & 0.48 \\
         C$_2$H$_{6}$ & 4,000K & 320 & 0.075 & 0.021 \\        
    \end{tabular}
    \caption{ The Wasserstein $W_1$ distance between predicted and extracted from MD simulations size distributions for molecules containing up to 20 carbons. This table shows how the accuracy of the prediction depends on the size of the system.}
    \label{table:size_effect}
\end{table*}

\begin{figure*}[!htb]
  \centerline{
 (a) \includegraphics[width=\textwidth]{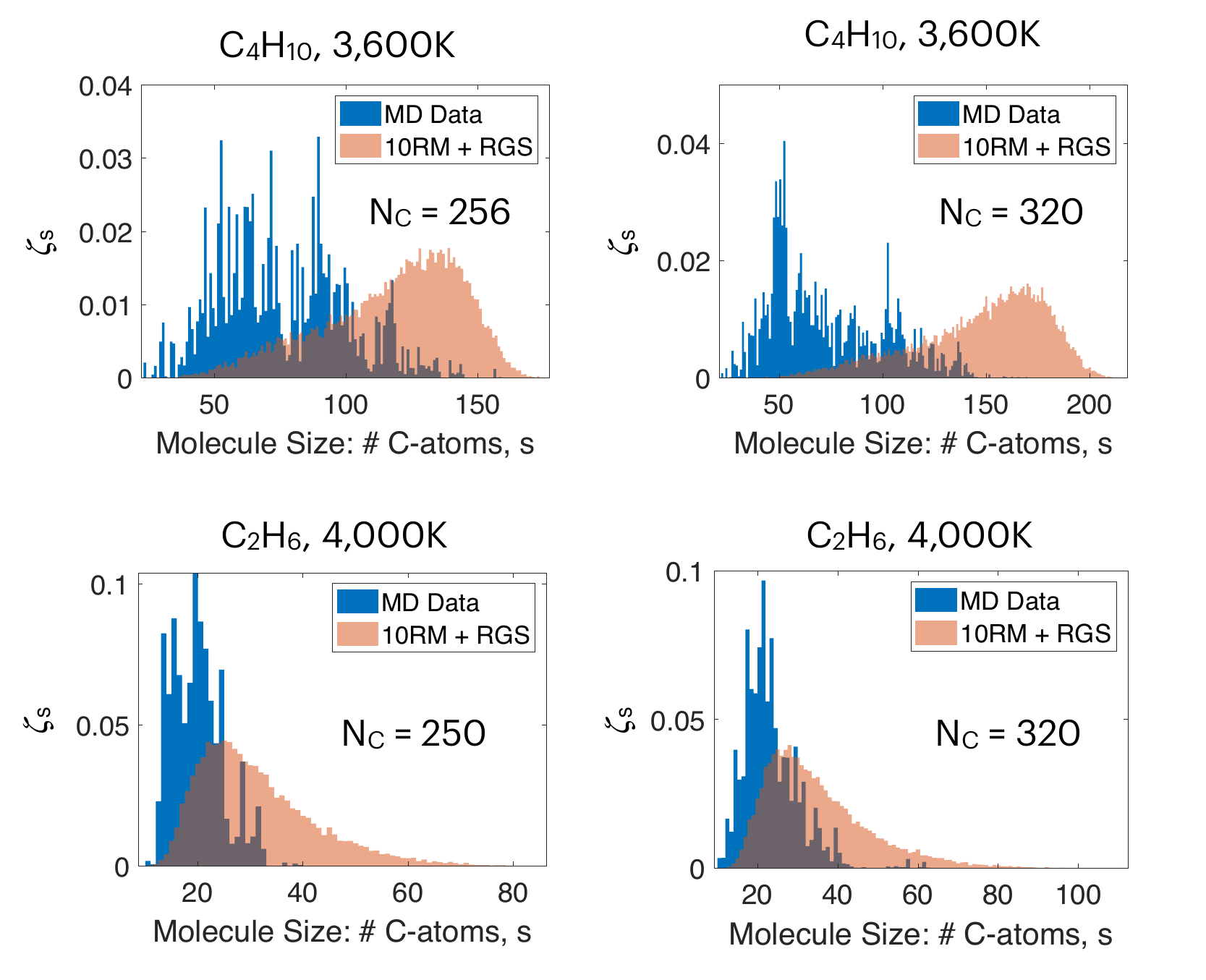}
 }
  \caption{Size distributions for the largest molecule predicted by 10RM+RGS and extracted from MD simulations. This figure illustrates the accuracy of the predictions for the molecule size distribution on the size of the system.
  } 
  \label{fig:size_effect1}
\end{figure*}

On the other hand, since the number of carbon atoms is one of the input parameters for the random graph sampling (RGS), we do not expect the accuracy of predictions by RGS to be affected by the system size. This expectation is reconfirmed by the side-by-side comparison of the histograms for the largest molecule size distribution predicted by 10RM+RGS and extracted from MD simulations -- see Fig. \ref{fig:size_effect1} and Table \ref{table:size_effect1}. Normalizing the data in column `RGS' by the number of carbons, we see that the size distribution for the largest molecule properly scales with $N_C$:
\begin{align*}
{\rm C}_4{\rm H}_{10}:&~~\frac{118\pm 26}{256} \approx \frac{147\pm 31}{320} 
\approx  0.46\pm 0.10,\\
{\rm C}_2{\rm H}_{6}:&~~\frac{32\pm 11}{250} \approx 0.13\pm 0.04,
\quad\frac{35\pm 13}{320} \approx 0.11\pm 0.04.
\end{align*}
On the other hand, the predictions for the largest molecule size distribution extracted from MD simulations are too noisy to determine how it scales with size.
\begin{table}[!htb]
    \centering
    \begin{tabular}{c|c|c|c|c}
        Init. Comp. & Temperature & \# carbons, $N_C$ & RGS &  MD \\
         \hline
         C$_4$H$_{10}$ & 3,600K & 256 & $118\pm 26$ & $77\pm 24$ \\
         C$_4$H$_{10}$ & 3,600K & 320 & $ 147\pm 31$ &$73\pm 28$ \\
         \hline
         C$_2$H$_{6}$ & 4,000K & 250 & $32\pm11$ & $20\pm 5$ \\
         C$_2$H$_{6}$ & 4,000K & 320 & $35\pm13$ & $24\pm 7$ \\        
    \end{tabular}
    \caption{The Wasserstein $W_1$ distance between predicted and extracted from MD simulations size distributions for molecules containing up to 20 carbons. This table shows how the accuracy of the prediction depends on the size of the system.}
    \label{table:size_effect1}
\end{table}

\pagebreak

\subsection{Prediction of histograms for particular molecule sizes} 
\label{sec:hist_small}

 \begin{figure*}[!htb]
  \centerline{
 (a) \includegraphics[width=0.7\textwidth]{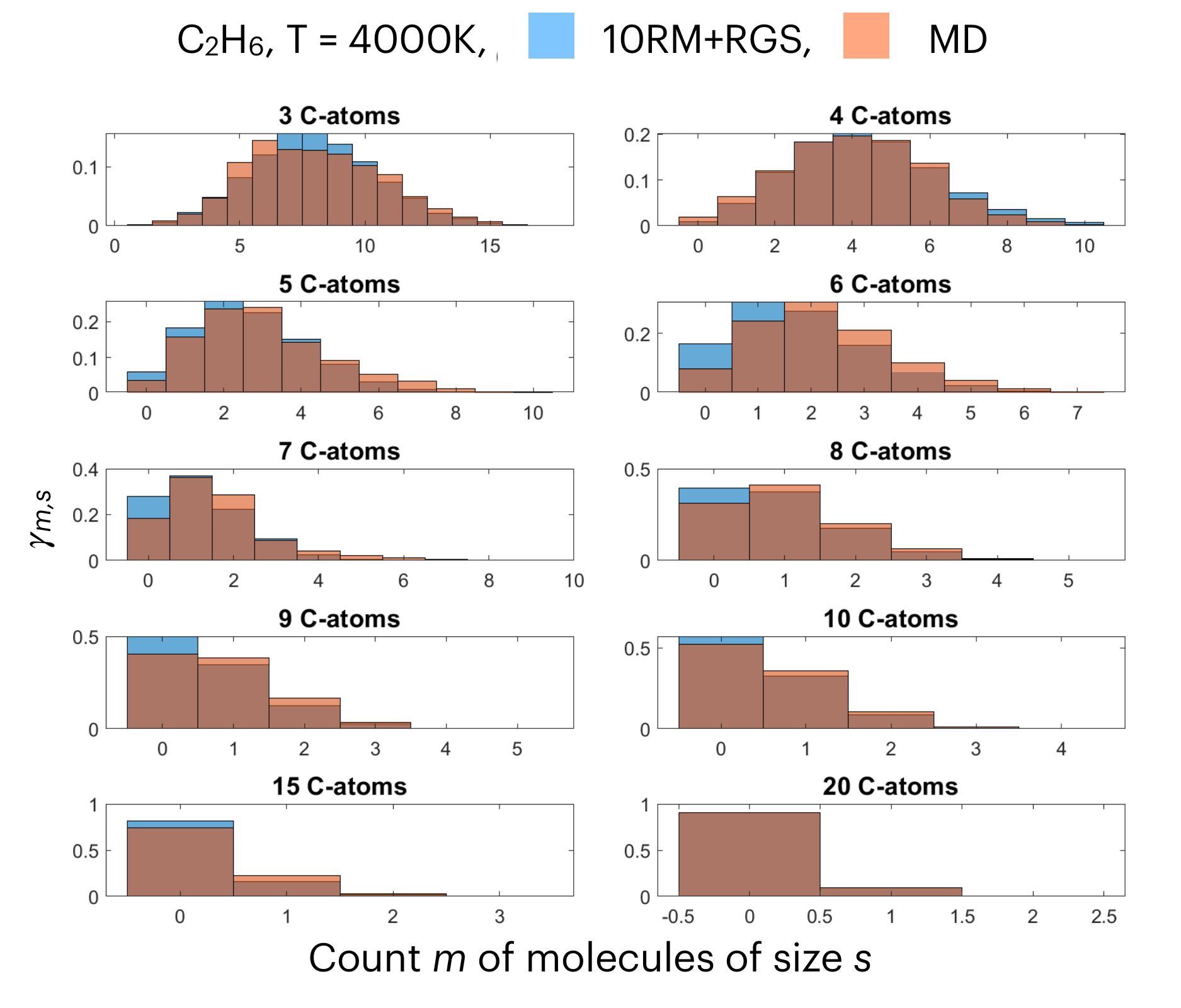}
 }
  \centerline{
 (b)\includegraphics[width=0.7\textwidth]{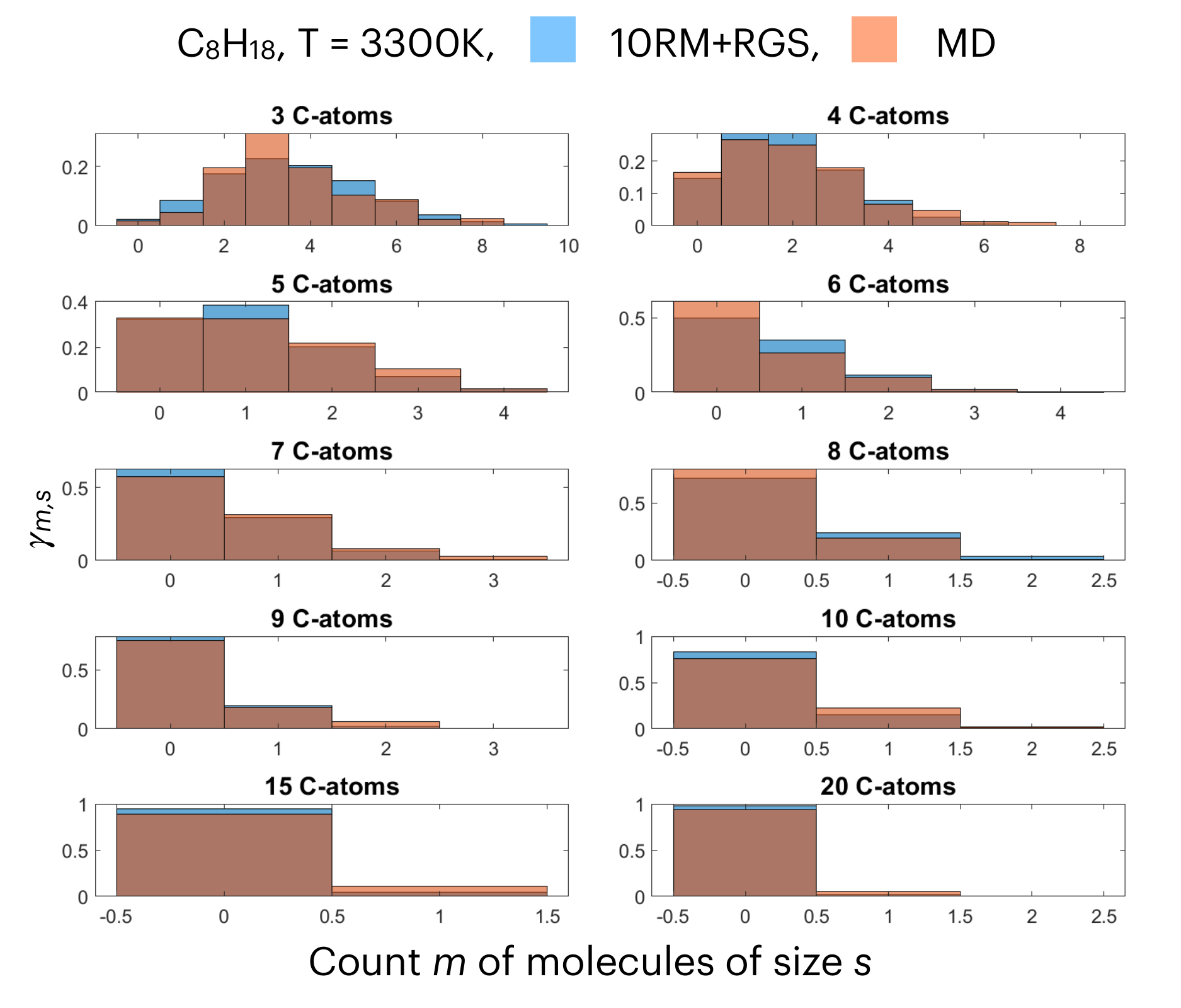}
 }
  \caption{Probability distributions for counts of small molecules of particular size $s$ obtained from random graph sampling using degree distribution from the ten-reaction model (``10RM + RGS'') (blue) and MD simulations (red). 
  The $x$-axis of each histogram is number $m$ of connected components or molecules containing the given number $s$ of carbon atoms. The $y$-axis is the probability $\gamma_m$ to have $m$ molecules of size $s$.  %
  } 
  \label{fig:SmallMolHist}
\end{figure*}

In this section, we will focus on a different class of distributions for small molecules. Specifically, given a molecule size $s>1$, what is probability $\gamma_{m,s}$ that there are $m$ molecules of size $s$ in the system? 

For each initial composition and temperature listed in the beginning of Table \ref{table:MD},
we generated $10,000$ random graphs using random graph sampling (``RGS'') as described in Section III C. The degree distribution was predicted by the ten-reaction model in each case. This method will be referred to as the ``10RM + RGS'' model. For each molecule size $1<s\le 20$, we generated histograms for the probabilities $\gamma_{m,s}$ to observe $m$ molecules of size $s$ and compared them to the histograms for the same probabilities extracted from the MD simulations. Two of these histogram comparisons are displayed in Fig. \ref{fig:SmallMolHist}.

\begin{figure*}[!htb]
 \centerline{
(a) \includegraphics[width =0.4\textwidth]{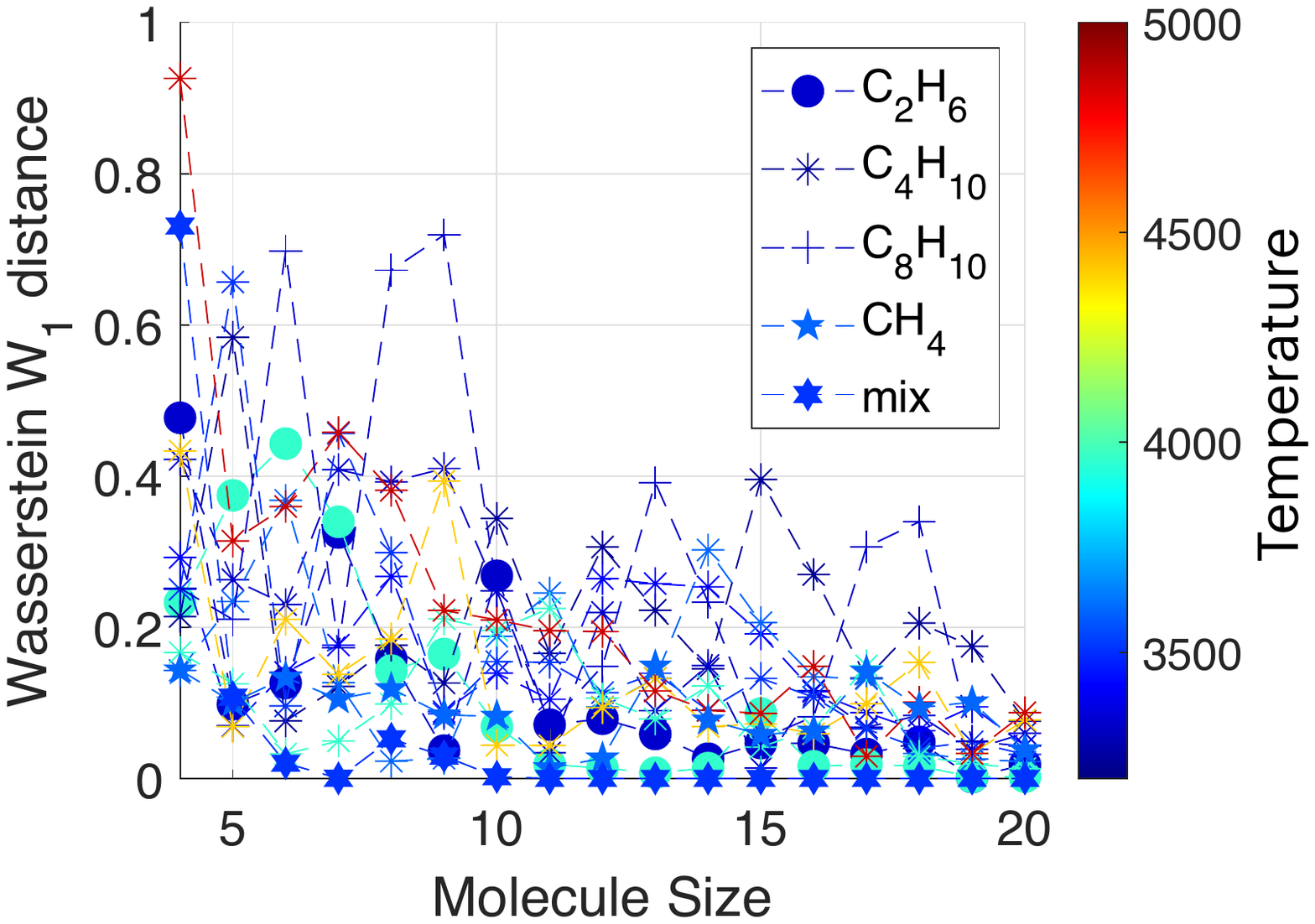}
(b) \includegraphics[width = 0.4\textwidth]{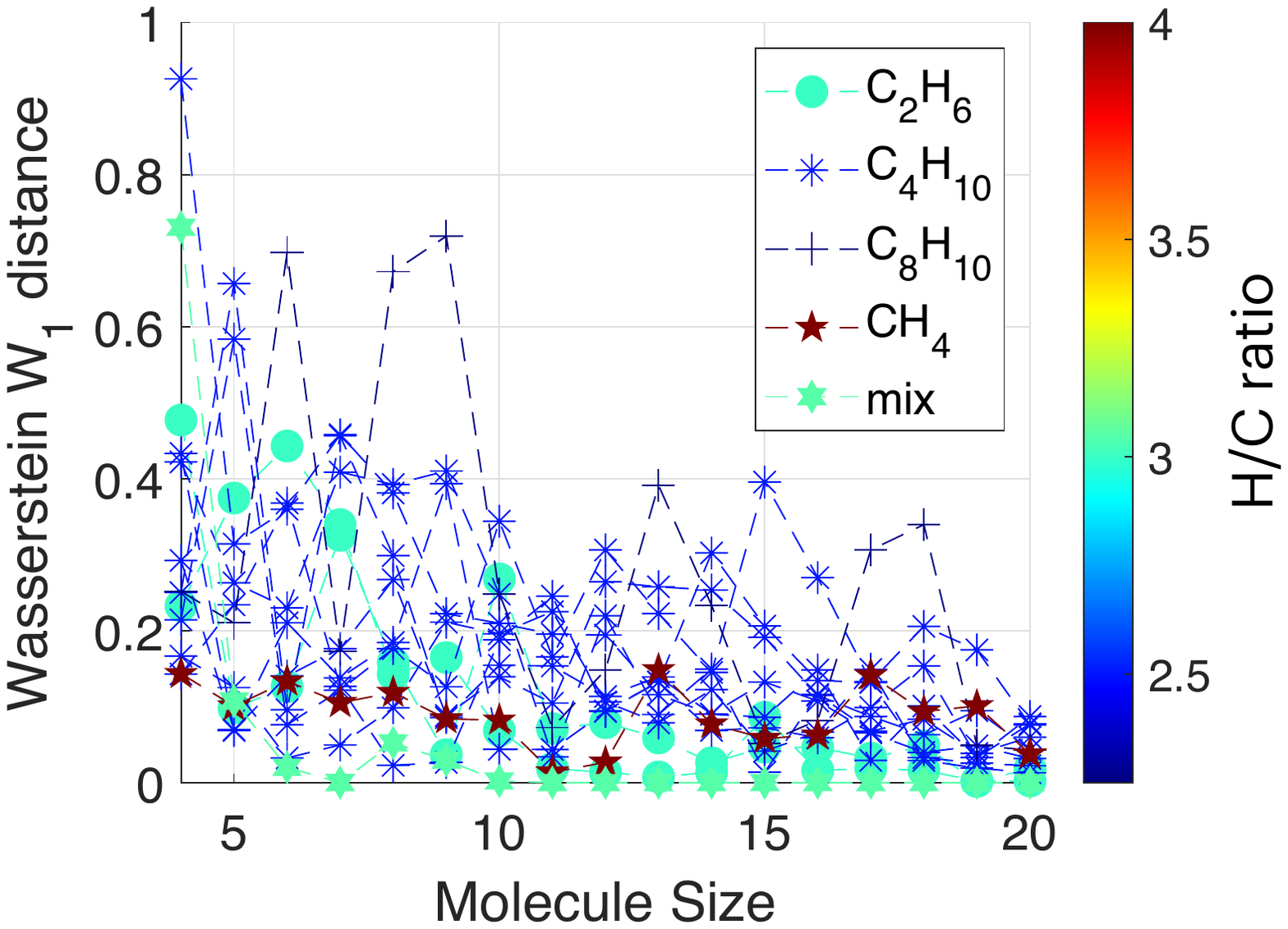}
}
 \caption{The Wasserstein-1 distance $W_1(P,Q)$ where $P$ is the molecule size distribution extracted from MD simulation and $Q$ is the predicted molecule size distribution obtained from the 10RM + RGS model as a function of $N$ for small molecule histograms. In (a) the data are color coded with system temperature and in (b) they are color coded with H/C ratio. Comparing figures (a) and (b) we see that the Wasserstein-1 distances between the predicted and the actual distributions tend to be smaller for higher H/C ratio while there is no significant dependence on temperature.}
  \label{fig:W1smallmoll}
\end{figure*}

We have found a very good agreement between the probability distributions $\gamma_{m,s}$ obtained using the 10RM + RGS model and extracted from the MD simulations. Fig. \ref{fig:W1smallmoll} quantifies the discrepancies between these distributions by means of the Wasserstein $W_1$ distance plotted as a function of molecule size. Figs. \ref{fig:W1smallmoll} (a) and (b) feature color coding of data points according to the temperature and the H/C ratio, respectively. We observe reasonably small Wasserstein distances  with one notable exception being the case where the molecule size $s=1$. The large discrepancy for $s=1$ is an artifact of our random graph sampling technique which always generates the same number of vertices of degree zero whereas in the MD simulations the number of methane molecules will vary slightly in time. 

\section{Acknowledgements}
This work was partially supported by the Air Force Office of Scientific Research under award number FA9550-20-1-0397.

\clearpage
\bibliography{bibliography}

\end{document}